\documentclass[journal]{IEEEtran}

\usepackage[noadjust]{cite}
\usepackage{graphicx}
\usepackage{float}
\usepackage{amsmath}
\usepackage{enumerate}
\usepackage{amssymb}
\usepackage{fixltx2e}
\usepackage{color}
\usepackage{algorithm}
\usepackage{algorithmic}
\usepackage{multirow}
\usepackage{ragged2e}
\usepackage{bm,flushend}

\allowdisplaybreaks[4]
\usepackage{mathtools}
\begin{document}


\title{Cooperative Backscatter Communications with Reconfigurable Intelligent Surfaces: An APSK Approach}

\author{Qiang~Li,~\IEEEmembership{{Member,~IEEE}}, Yehuai~Feng, Miaowen Wen,~\IEEEmembership{{Senior Member,~IEEE}}, Jinming Wen, \\George C. Alexandropoulos,~\IEEEmembership{{Senior Member,~IEEE}}, Ertugrul Basar,~\IEEEmembership{{Fellow,~IEEE}}, \\and H. Vincent Poor,~\IEEEmembership{{Life Fellow,~IEEE}} \\
	   \thanks{Qiang~Li, Yehuai~Feng, and Jinming Wen are with the College of Information Science and Technology, Jinan University, Guangzhou 510632, China (e-mail: qiangli@jnu.edu.cn; fengyh@stu2022.jnu.edu.cn; jinming.wen@mail.mcgill.ca).}
	   \thanks{Miaowen Wen is with the School of Electronics and Information Engineering, South China University of Technology, Guangzhou 510640, China (e-mail: eemwwen@scut.edu.cn).}
	   \thanks{George C. Alexandropoulos is with the Department of Informatics and Telecommunications, National and Kapodistrian University of Athens, 15784 Athens, Greece (e-mail: alexandg@di.uoa.gr).}
	   \thanks{Ertugrul Basar is with the Department of Electrical and Electronics Engineering, Ko\c{c} University, Istanbul 34450, Turkey (e-mail: ebasar@ku.edu.tr).}
	   \thanks{H. Vincent Poor is with the Department of Electrical and Computer Engineering, Princeton University, Princeton, NJ 08544, USA (e-mail: poor@princeton.edu).}
}

\maketitle

\begin{abstract}
In this paper, a novel amplitude phase shift keying (APSK) modulation scheme for cooperative backscatter communications aided by a reconfigurable intelligent surface (RIS-CBC) is presented, according to which the RIS is configured to modulate backscatter information onto unmodulated or PSK-modulated signals impinging on its surface via APSK. We consider both passive and active RISs, with the latter including an amplification unit at each reflecting element. In the passive (resp. active) RIS-CBC-APSK, backscatter information is conveyed through the number of RIS reflecting elements being on the ON state (resp. active mode) and their phase shift values. By using the optimal APSK constellation to ensure that reflected signals from the RIS undergo APSK modulation, a bit-mapping mechanism is presented. Assuming maximum-likelihood detection, we also present closed-form upper bounds for the symbol error rate (SER) performance for both passive and active RIS-CBC-APSK schemes over Rician fading channels. In addition, we devise a low-complexity detector that can achieve flexible trade-offs between performance and complexity. Finally, we extend RIS-CBC-APSK to multiple-input single-output scenarios and present an alternating optimization approach for the joint design of transmit beamforming and RIS reflection. Our extensive simulation results on the SER performance corroborate our conducted performance analysis and showcase the superiority of the proposed RIS-CBC-APSK schemes over the state-of-the-art RIS-CBC benchmarks.
\end{abstract}

\begin{keywords}
Reconfigurable intelligent surface (RIS), amplitude phase shift keying (APSK), symbol error rate (SER), performance analysis, cooperative backscatter communications.
\end{keywords}

\IEEEpeerreviewmaketitle

\section{Introduction}
\subsection{Background and Related Work}
Cooperative backscatter communications (CBC) enable a backscatter transmitter (B-Tx) to convey information by modulating and reflecting radio-frequency (RF) signals received from an active transmitter (A-Tx), and a cooperative receiver (C-Rx) is deployed to jointly decode the information received from the B-Tx and the A-Tx \cite{YangCooperative,LiangSymbiotic}. In CBC, the B-Tx embeds information into incident modulated signals via backscatter modulation, which dispenses with complex RF chains and enjoys low power consumption. Moreover, the information communication via the A-Tx can benefit from the backscatter transmission of the same by the B-Tx, considering that the backscatter link acts as an additional multipath component towards the C-Rx. Therefore, CBC achieves a win-win effect for the active and backscatter transmission. In conventional CBC systems, a B-Tx often employs a single or a few reflecting antennas to enable the functionalities of information modulation and reflection. However, since the backscatter transmission experiences ``multiplicative fading'', the signal power of the backscatter link tends to be weak, which limits the performance of CBC. 

Reconfigurable intelligent surfaces (RISs) have recently attracted significant interest from both academia and industry due to their capability to reshape the radio environment \cite{YLiu_RIS_Principles_and_Opportunities,Renzo_smart,BasarEmerging}. An RIS is a panel consisting of numerous basic unit elements, whose amplitude and phase responses on impinging electromagnetic signals can be dynamically programmed. All RIS elements can reflect incident signals by exerting adjustable reflection amplitudes and phase shifts with negligible power consumption and noise interference \cite{Yang_dual_frequency}. For this prominent property, RISs have been recently considered for various wireless applications and objectives, such as improving error performance \cite{LiChannel}, energy efficiency \cite{Huang_energy_efficiency}, physical-layer security \cite{Alexandropoulos_Counteracting}, network coverage \cite{Alexandropoulos_smart}, integrated sensing and communications \cite{Chepuri_ISCR}, and spatiotemporal signal focusing \cite{Alexandropoulos_propagation}. Thus, equipping a B-Tx with an RIS is expected to be an effective solution to the problem of ``multiplicative fading'' in CBC. This advanced paradigm of CBC, termed hereinafter as RIS-CBC, is considered to be a promising solution for future low-power Internet-of-Things \cite{Liang_RIS_BC,Lei_RIS_SR}.

A key problem in RIS-CBC is to design RIS-based modulation at the B-Tx to convey backscatter information. In \cite{MaReconfigurable}, an RIS-coated B-Tx transmitted a binary phase shift keying (BPSK) symbol by changing the polarity of RIS beamforming every several active symbols. By recalling that RISs can be configured at a symbol rate \cite{TangWireless}, the transmission rate of B-Tx can be improved by configuring the RIS for every active symbol. In \cite{Li_SSK} and \cite{Singh_RIS}, RIS-based Alamouti coding and PSK were conceived to convey backscatter information while reflecting incident space shift keying signals to a C-Rx, by deploying the RIS as two transmit antennas and one transmit antenna, respectively. In addition, the concept of spatial modulation (SM) \cite{WenA} can be applied to an RIS, which enables the RIS to carry backscatter information by indexing RIS elements. Specifically, in \cite{Yan_onoff_RIS}, the ON/OFF state of each RIS element was exploited to carry one bit of information. In \cite{LinReconfigurable} and \cite{Hussein_RM}, the indices of the RIS elements in the ON state were used to carry information, while the number of active elements was fixed for each transmission. To overcome the power loss resulting from inactive elements, the authors in \cite{Lin_RIS_QRM} proposed to activate some RIS elements into an in-phase (I) state and the rest into a quadrature (Q) state, where backscatter information was conveyed by the indices of the elements in the I state. In contrast to \cite{Lin_RIS_QRM}, ref. \cite{Li_NM} proposed to employ the number of RIS elements in the I state to carry backscatter information, resulting in a scheme known as RIS-aided number modulation (RIS-NM). In \cite{Basar_RIS_RSM} and \cite{Ma_RIS_TRSM}, considering an C-Rx is equipped with multiple antennas, backscatter information was conveyed by an antenna index of the C-Rx per transmission, during which an RIS was configured to maximize the receive signal-to-noise ratio (SNR) at that antenna.
In \cite{Li_SSK}-\cite{Ma_RIS_TRSM}, the active and backscatter information was transmitted by the two different information-carrying units of SM. Note that this information can also be transmitted via a composite two-dimensional constellation. In \cite{Yang_RIS_phaseshift_JAPT}, an RIS was presented to superimpose irregular BPSK symbols on $M$-ary PSK symbols transmitted from the A-Tx, and thus, a regular $2M$-PSK constellation was constructed from the active and backscatter information. By recalling that amplitude PSK (APSK) tends to achieve better error performance than PSK in the case of high modulation orders, using APSK is expected to be a promising way to convey active and backscatter information. However, related work in this field only includes \cite{Wu_RIS_star_QAM}, which proposes to map partial backscatter information into the index of an Rx antenna, and modulate the remaining backscatter information and the PSK-modulated active symbols into a double-ring star-quadrature amplitude modulation (QAM) symbol.

In the above-mentioned RIS-CBC schemes, passive RISs were used, i.e., the RISs completely consist of passive reflecting elements with negligible power consumption and without incurring thermal noise. One drawback of passive RISs, however, is that they reflect incident signals without amplifying them \cite{BasarEmerging}, and thus a large number of RIS elements are required to combat the resulting ``multiplicative fading'' effect. Recently, active RISs have emerged as an alternative to passive RISs \cite{ZhangActiveRIS,Tasci_single_power_amplifier}. In contrast to passive RISs, active RISs are able to amplify incident signals by integrating a reflection-type amplifier into each \cite{ZhangActiveRIS} or group \cite{Tasci_single_power_amplifier} of the reflecting elements. However, active RISs consume more power and introduce non-negligible thermal noise, compared with passive RISs. It is shown in \cite{Zhi_active_RIS_power_budget} that, with the same small power budget, passive RISs perform better than active RISs, while the situation is reversed for a sufficient power budget. Very recently, active RISs have been considered for CBC. In \cite{Sanila_ARIS_JSRM}, an A-Tx employs SM, while a B-Tx embeds backscatter information into the indices of the element groups in the ON state and a targeted receive antenna. In RIS-aided hybrid reflection modulation (RIS-HRM), based on an active RIS that consists of both active and passive reflecting elements, a B-Tx implements a virtual amplitude shift keying (ASK) modulator by adjusting the number of active elements \cite{Yigit_RIS_HRM}.

\subsection{Motivation and Contribution}
As discussed in the previous subsection, at the B-Tx of RIS-CBC, a composite two-dimensional constellation can be generated to simultaneously convey the active and backscatter information. By considering that the constellation symbol should carry both the active and backscatter information, the modulation order tends to be high. In this case, APSK is likely to be a better choice than PSK and ASK. However, in the existing RIS-CBC with APSK \cite{Wu_RIS_star_QAM}, the number of constellation rings is limited to two, and each ring includes the same number of constellation points. Thus, the considered APSK constellation is far from being optimal. Further, since the APSK is carried out by adjusting both the phase shifts and reflection amplitudes of the RIS, the RIS configuration is complicated. Finally, \cite{Wu_RIS_star_QAM} only considers the case of passive RISs, and the combination of CBC with active RISs and APSK has not yet been considered in the open technical literature. 

Against the latter background, an RIS-CBC scheme, termed as RIS-CBC-APSK, is proposed in this paper, which implements a generalized RIS-based APSK constellation with optimized Euclidean distances to modulate backscatter information onto PSK-modulated incident signals. RIS-CBC-APSK applies to both passive and active RISs. The main contributions of this paper are summarized as follows:

\begin{itemize}
	\item An RIS-based APSK scheme is designed for CBC, in which the amplitude modulation is achieved by turning ON/OFF the RIS reflecting elements and reflection amplifiers for the cases of passive and active RISs, respectively, and the phase modulation is implemented by adjusting the RIS phase shifts. A bit-mapping mechanism at the RIS is developed with the aim to maximize the minimum Euclidean distance between two distinct constellation points. RIS-CBC-APSK subsumes the special case where an A-Tx transmits unmodulated carrier waves.

	\item We derive closed-form upper bounds on the symbol error rate (SER) of the active and backscatter information for both active and passive RIS-CBC-ASPK with maximum likelihood (ML) detection over Rician fading channels. A low-complexity (LC) detector that can achieve flexible trade-offs between performance and complexity is developed for RIS-CBC-ASPK. Further, we extend RIS-CBC-APSK to multiple-input single-output (MISO) scenarios by optimizing the beamforming at the A-Tx and RIS alternately.
	
	\item Extensive computer simulation results are presented. They showcase that the proposed RIS-CBC-APSK scheme significantly outperforms existing RIS-CBC ones in terms of SER, especially for  backscatter transmission. Moreover, in the case of unmodulated incident signals, RIS-CBC-APSK exhibits superiority over existing RIS-CBC schemes. The presented SER analysis, LC detector, and extension to MISO are also verified by  simulations.  
\end{itemize}

The remainder of this paper is organized as follows. The optimization of APSK constellations is revisited in Section~II. Then, Sections~III and IV describe the system model and analyze the error performance for passive and active RIS-CBC-APSK, respectively. Further, we design a LC detector and extend RIS-CBC-APSK to MISO systems in Section V. Section VI presents computer simulation results, and finally, Section~VII concludes the paper.

\textit{Notation:} Column vectors and matrices are denoted by lowercase and uppercase boldface letters, respectively. Superscripts $(\cdot)^*$, $(\cdot)^T$, and $(\cdot)^H$ stand for conjugate, transpose, and Hermitian transpose, respectively. $X _{{\imath }{\jmath }}$ denotes the ($\imath,\jmath$)-th element of $\mathbf{X}$ and $x_\imath$ denotes the $\imath$-th element of $\mathbf{x}$. $|x|$, $\angle(x)$, and $\Re\{x\}$ return the amplitude, phase, and real part of $x$, respectively. $j=\sqrt{-1}$ is the imaginary unit. $(\mathcal{C})\mathcal{N}(\mu,\sigma^2)$ represents the (complex) Gaussian distribution with mean $\mu$ and variance $\sigma^2$. $ \mathbf{I}_{n} $ denotes the identity matrix of size $n \times n$. $\mathbb{C} ^{n\times m}$ denotes the set of $n \times m$ complex-valued matrices. $\mathrm{diag}\{\cdot\}$ transforms a vector into a diagonal matrix. $\mathrm{trace}(\mathbf{X})$ and $\mathrm{rank}(\mathbf{X})$ denote the trace and rank of $\mathbf{X}$, respectively. The probability of an event is denoted by $\Pr(\cdot)$. $\mathrm{E}\{\cdot\}$ and $\mathrm{VAR}\{\cdot\}$ denote the  expectation and variance, respectively. $Q(\cdot)$, $\varGamma(\cdot)$, and ${}_{1}F_1(\cdot;\cdot;\cdot)$ represent the Gaussian $Q$-function, the gamma function, and the confluent hypergeometric function \cite{SimonProbability}, respectively. $\left\| \cdot \right\|$ denotes the Frobenius norm. $\mathbb{I}(\cdot)$ and $\lfloor\cdot\rceil$ represent the indicator function and the rounding off operation, respectively. $\mathbf{X}\succeq  0$ denotes that $\mathbf{X}$ is a positive semi-definite matrix. $\text{sort}(\cdot,\uparrow)$ sorts a vector in ascending order.

\section{Preliminaries on APSK Constellations }
\label{APSK Constellation Revisited}
\begin{figure}[t]
	\centering
	\includegraphics[width=3.3in]{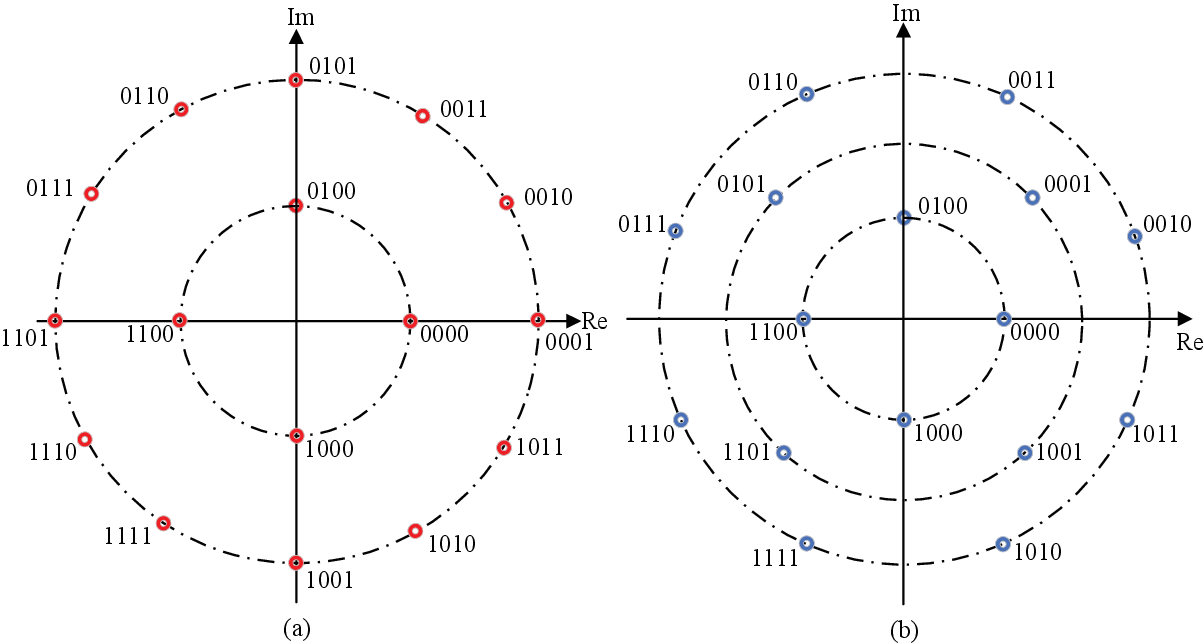}
	\caption{Examples of $16$-APSK constellations: (a) $4+12$-APSK and (b) $4+4+8$-APSK.}
	\label{Constellation_M16}
\end{figure}
An APSK constellation is composed of multiple concentric rings, each with uniformly spaced PSK points. For an $M$-ary APSK constellation with $R$ rings, denoted by $\mathcal{S}$, each $\imath$-th constellation point on each $k$-th ring is given by
\begin{align}\label{APSK_constellation}
	s = r_k\exp \left( j\left( \frac{2\pi}{n_k}\imath+\vartheta _k \right) \right), 
\end{align}
where $r_k$, $n_k$, and $\vartheta_k$ are the radius, the number of points, and the initial phase offset of the $k$-th ring, respectively, with $\imath \in \{0,\ldots,n_k-1\}$, $k \in \{1,\ldots,R\}$, and $n_1+n_2+\dots+n_R=M$. It is obvious that, for a given value of $M$, there are multiple choices of $R$ and $n_k$. Hence, to remove ambiguity, we usually refer to $M$-APSK as $n_1+n_2+\dots+n_R$-APSK specifically. For example, with $M=16$, both $4+12$-APSK and $4+4+8$-APSK are possible, which are shown in Figs.~\ref{Constellation_M16}(a) and (b), respectively. It should be noted that there are two possibilities for the normalization of APSK symbols, i.e., $\mathrm{E}\{|s|^2\}=1$ and $r_R=1$ \cite{FSCCTC}. Without loss of generality, we adopt the normalization $r_R=1$ in this paper.

It can be observed from (\ref{APSK_constellation}) that the construction of an $M$-APSK constellation includes three steps: selecting $R$ and $n_k$, determining $r_k$, and choosing $\vartheta_k$. In general, the values of $R$ and $n_k$ are selected under the constraints of $n_1+n_2+\dots+n_R=M$ and $n_k\leq n_{k+1}$ \cite{ACT:Opti-MAPSK}. For given $R$ and $n_k$, the determination of $r_k$ and $\vartheta_k$ is an interesting optimization problem. A frequently used objective is to maximize the minimum squared Euclidean distance, namely:
\begin{align}
	\text{(P1)}: \underset{r_1,\ldots,r_{R-1},
		\atop \vartheta_{1},\ldots,\vartheta_{R}}{\text{max}}\quad&~  \min _{s,\hat{s}\in \mathcal{S}, s\ne\hat{s} }| s-\hat{s} |^2 \label{P1} \\
	s.t. \quad&~
	r_R=1\tag{\ref{P1}{a}} \label{P1a},\\
	&~
	0\leq \vartheta _k< 2\pi, \ k=1,\ldots ,R\tag{\ref{P1}{b}} \label{P1b},\\
	&~
	r_1<\cdots<r_R\tag{\ref{P1}{c}} \label{P1c}.
\end{align}

For high-order APSK, $\vartheta_k$ for all $k$ have less influence on the minimum squared Euclidean distance \cite{IJSCN:Turbo-APSK,SPSC:64MAPSK}. Hence, we can set $\vartheta_1=\cdots=\vartheta_R=0$ for simplicity. Moreover, thanks to the symmetry of APSK constellations, to calculate the minimum squared Euclidean distance, we only need to focus on the adjacent points in a ring and the nearest points between adjacent rings, which can be expressed as follows:
\begin{align}\label{d_k2}
	d_{k}^{2}=2r_{k}^{2}\left( 1-\cos \frac{2\pi}{n_k} \right) 
\end{align}
and
\begin{align}\label{d_k_k12}
	d_{k,k+1}^{2}=(\gamma_k-1)^2r_{k}^{2},
\end{align}
respectively, with $\gamma _k=r_{k+1}/r_k$. Therefore, problem (P1) can be simplified as
\begin{align}
	\text{(P1.1)}: 	\underset{ \gamma_{1},\ldots,\gamma_{R-1}}{\text{max}}\ &~  \min \left\{ d_{1}^{2},\ldots ,d_{R}^{2},d_{1,2}^{2},\ldots ,d_{R-1,R}^{2} \right\}\label{P1.1} \\
	s.t. \quad&~
	r_R=1\tag{\ref{P1.1}{a}} \label{P1.1a},\\ 
	&~
	1<\gamma_k<\varUpsilon ,\ k=1,\ldots ,R-1\tag{\ref{P1.1}{b}} \label{P1.1b},
\end{align}
where $\varUpsilon$ is a heuristic upper bound on $\gamma_k$ that can be conservatively set to 4 \cite{ACT:Opti-MAPSK,FSCCTC,IJSCN:Turbo-APSK,SPSC:64MAPSK}. Finally, problem (P1.1) can be solved via exhaustive search. 

\section{Proposed Passive RIS-CBC with APSK}
In this section, we present a passive RIS-CBC-APSK scheme based on the APSK constellations derived in Section~\ref{APSK Constellation Revisited}, and then analyze its error performance by assuming ML detection.
\subsection{System Model}
\begin{figure}[!tb]
	\centering
	\includegraphics[width=3.3in]{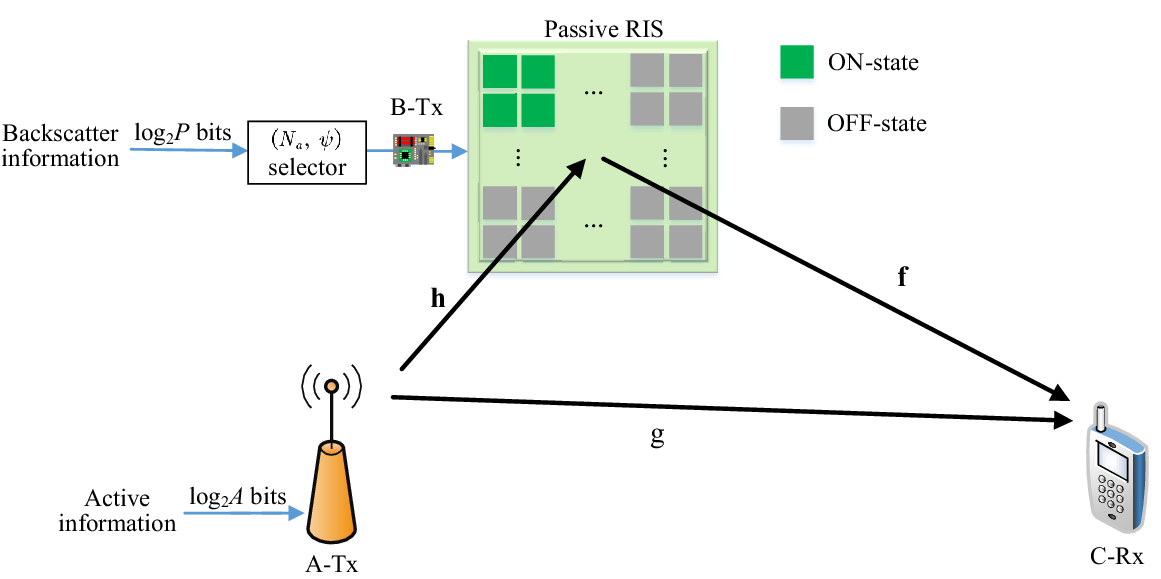}
	\caption{Block diagram of passive RIS-CBC-APSK in a SISO setting.}
	\label{SystemModel_1}
\end{figure}
A block diagram of the proposed passive RIS-CBC-APSK system with single-input single-output (SISO) transceivers is depicted in Fig.~\ref{SystemModel_1}, where a single-antenna A-Tx transmits active information via $A$-ary PSK to a single-antenna C-Rx through a direct and an RIS-aided links. In addition, backscatter information is conveyed from the passive RIS through the electromagnetic waves transmitted from the A-Tx that imping on it and get reflected. The RIS, positioned in the proximity of the A-Tx, is configured by an RIS controller that receives instructions from the A-Tx via a dedicated link \cite{BasarEmerging}. The channel links from the A-Tx to the RIS and from the RIS to the C-Rx are represented by $\mathbf{h}=[ h_1,\ldots ,h_N ] ^T$ and $\mathbf{f}=[ f_1,\ldots ,f_N ] ^T$, respectively, where $h_i$ ($f_i$) denotes the channel coefficient from the A-Tx ($i$-th element of the RIS) to the $i$-th element of the RIS (C-Rx) with $i=1,\ldots,N$ (i.e., $N$ RIS elements in total). The direct link from the A-Tx to the C-Rx is denoted by $g$. The reflection matrix of the passive RIS is modeled as $\mathbf{\Phi}=\mathrm{diag}\{[ \beta_1 e^{ j\phi_1 },\ldots,\beta_N e^{j\phi_N}]^T\}$, where $\beta_i\in \{0,1\}$ and $\phi_i \in [0,2\pi)$ are the amplitude coefficient and phase shift of the $i$-th RIS element, respectively.

The received signal at the C-Rx can be expressed as follows \cite{Huang_energy_efficiency}:
\begin{align}\label{received_y}
	y=\sqrt{P_t}\left( \mathbf{f}^T\mathbf{\Phi}\mathbf{h}+g \right) x+w,
\end{align}
where $P_t$ denotes the transmit power of the A-Tx, $w$ is a sample of additive white Gaussian noise with the distribution $\mathcal{CN}(0,N_0)$, and $x$ is an $A$-ary PSK symbol drawn from a constellation set $\mathcal{A}$ with normalized average power. The backscatter information is conveyed by $\mathbf{\Phi}$, which is designed as follows. To enable APSK at the RIS, not all RIS elements are turned ON. Moreover, we adjust the number of the RIS elements with ON-state (or OFF-state, alternatively) and the associated phase shifts according to the backscatter information to be transmitted. Here, for the elements with ON-state and OFF-state, the amplitude coefficients are set to $1$ and $0$, respectively. With the first $N_a$ RIS elements turned ON, (\ref{received_y}) is rewritten as
\begin{align}\label{received_y1}
	y=\sqrt{P_t}\left(\sum_{i=1}^{N_a}f_ih_i e^{j\phi _i}  + g \right) x + w.
\end{align}
To maximize the received power at the C-Rx and to enable backscatter transmission, $\phi_i$ is set to\footnote{Since there are plenty of channel estimation schemes in literature for RIS-aided communications (e.g., \cite{Jian_est} and \cite{Swindlehurst_channel_esti}), perfect CSI is assumed at the RIS in this paper.}
\begin{align}\label{thata_value}
	\phi _i=-\angle \left( f_ih_i \right) +\angle g+\psi.
\end{align}
Then, by defining $\underline{h}_i = | f_i || h_i |$ and $\mathcal{H} _{N_a}=\sum\nolimits_{i=1}^{N_a}{\underline{h}_i}$, we have that:
\begin{align}\label{receieved_y2}
	y=\sqrt{P_t}\left( e^{j\psi}\mathcal{H} _{N_a}+\left| g \right| \right) xe^{j\angle g}+w,
\end{align}
where $\psi$ and $N_a$ in (\ref{received_y1}) are used to carry backscatter information.

\begin{figure}[!tb]
	\centering
	\includegraphics[width=3.3in]{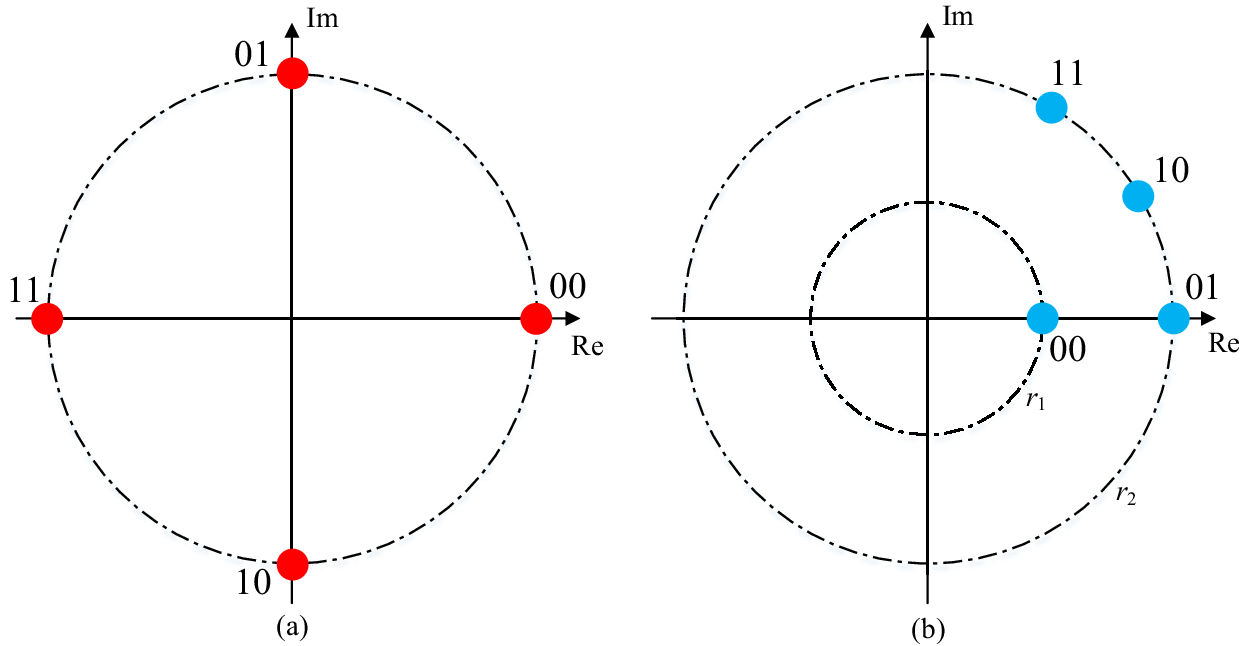}
	\caption{Constellations of (a) $x$ and (b) $\mathcal{H} _{N_a}e^{j\psi}$, for $A=P=4$.}
	\label{Constellations_AP}
\end{figure}

It is easy to observe from (\ref{receieved_y2}) that $\mathcal{H} _{N_a}xe^{j\psi}$ can constitute an APSK symbol $s$ with $|s|=\mathcal{H} _{N_a}$ and $\angle s = \angle ( xe^{j\psi} )$ by elaborately designing $N_a$ and $\psi$. After recalling $n_1+\cdots+n_R$-APSK constellations in Section \ref{APSK Constellation Revisited}, the set of all possible values of $N_a$ can be identified as 
\begin{align}\label{N_a}
	\mathcal{N}_a=\left\{ \lfloor N/\prod_{\varsigma=1}^{R-1}{\gamma _\varsigma}\rceil,\lfloor N/\prod_{\varsigma=2}^{R-1}{\gamma _\varsigma}\rceil, \ldots,\lfloor N/ \gamma _{R-1} \rceil,N\right\}, 
\end{align}
where the $k$-th entity of $\mathcal{N}_a$ corresponds to the symbols on the $k$-th ring of the $n_1+\cdots+n_R$-APSK constellation with $k=1,\ldots,R$. Moreover, when ${N}_a$ is taken as the $k$-th entity of $\mathcal{N}_a$, the set of all possible values of $\psi$ is given by $\{0,2\pi/n_k,\ldots,2\pi(n_k/A-1)/n_k\}$, where $n_k$ is a positive integer multiple of $A$. Therefore, there are $P=\sum_{k=1}^{R}n_k/A=M/A$ possible combinations of $(N_a,\psi)$, resulting in a total of $\log_{2}P$ bits of backscatter information. Hence, the data rate of RIS-CBC-APSK is $\log_2(AP)$ bits per channel use (bpcu). An example of the proposed bit mapping for passive RIS-CBC-APSK with $A=4$, $P=4$, and $4+12$-APSK is presented in Table \ref{Table 1}. The corresponding constellations of $x$ and $\mathcal{H} _{N_a}e^{j\psi}$ are shown in Figs.~\ref{Constellations_AP}(a) and (b), respectively, and the constituted APSK constellation is given in Fig.~\ref{Constellation_M16}(a). We note that the constituted APSK constellation is channel-dependent due to $\mathcal{H} _{N_a}$. Figs.~\ref{Constellation_M16}(a) and \ref{Constellations_AP}(b) show the cases with a realization of $\mathcal{H} _{N_a}$.  
\begin{table*}[!t]
	\caption{Bit mapping for RIS-CBC-APSK with $A=4$, $P=4$, and $4+12$-APSK.}
	\label{Table 1}
	\centering
	\begin{tabular}{|c|c|c||c|c|c|}
		\hline
		$\log_2A + \log_2P$ bits & $x$ 	& $(N_a,\psi)$ 					  &$\log_2A + \log_2P$ bits	& $x$ 	& $(N_a,\psi)$\\
		\hline
		\hline
		$[00|00]$ 				& $1$   & $(\lfloor N/\gamma _1\rceil,0)$ &$[10|00]$ 				&$-j$  	& $(\lfloor N/\gamma _1\rceil,0)$\\
		\hline
		$[00|01]$ 				& $1$	& $(N,0)$ 						  &$[10|01]$ 				&$-j$ 	& $(N,0)$\\
		\hline
		$[00|10]$ 				& $1$	& $(N,\pi/6)$ 					  &$[10|10]$				&$-j$ 	& $(N,\pi/6)$\\
		\hline
		$[00|11]$ 				& $1$	& $(N,\pi/3)$ 					  &$[10|11]$ 				&$-j$ 	& $(N,\pi/3)$ \\
		\hline
		$[01|00]$ 				& $j$	& $(\lfloor N/\gamma _1\rceil,0)$  &$[11|00]$ 				&$-1$ 	&$(\lfloor N/\gamma _1\rceil,0)$ \\
		\hline
		$[01|01]$ 				& $j$	& $(N,0)$ 					  	  &$[11|01]$ 				&$-1$ 	&$(N,0)$ \\
		\hline
		$[01|10]$ 				& $j$	& $(N,\pi/6)$ 					  &$[11|10]$ 				&$-1$ 	&$(N,\pi/6)$ \\
		\hline
		$[01|11]$ 				& $j$	& $(N,\pi/3)$ 					  &$[11|11]$ 				&$-1$ 	&$(N,\pi/3)$ \\
		\hline

	\end{tabular}

\end{table*}

At the C-Rx, the active and backscatter information can be recovered by estimating $x$ and $(N_a, \psi)$ jointly with ML detection, i.e.,
\begin{align}\label{ML_y}
	\left( \hat{N}_a ,\hat{\psi},\hat{x} \right) \!= \!\mathrm{arg} \min_{ N_a ,\psi,x } \left| y-\sqrt{P_t}\left(e^{j\psi} \mathcal{H} _{N_a}  +\left| g \right| \right) xe^{j\angle g} \right|^2,
\end{align}
where $\hat{N}_a$, $\hat{\psi}$, and $\hat{x}$ are the estimates of $N_a$, $\psi$, and $x$, respectively.

\newcounter{TempEqCnt} 
\setcounter{TempEqCnt}{\value{equation}} 
\setcounter{equation}{14} 
\begin{figure*}[ht]	
	\vspace*{1pt}
			\hrulefill
	\begin{align}\label{ConditionalPEP}
		&\mathrm{Pr}\left( \left( N_a, \psi,x \right) \rightarrow \left( \hat{N}_a,\hat{\psi},\hat{x} \right) |\left( \mathbf{h},\mathbf{f},g \right) \right) \nonumber
		\\
		&=\mathrm{Pr}\bigg( \left| y-\sqrt{P_t}\left( e^{j\psi}\mathcal{H} _{N_a}+\left| g \right| \right) xe^{j\angle g} \right|^2>\left| y-\sqrt{P_t}\left( e^{j\hat{\psi}}\mathcal{H} _{\hat{N}_a}+\left| g \right| \right) \hat{x}e^{j\angle g} \right|^2 \bigg) \nonumber \\
		&=\mathrm{Pr}\bigg( -P_t\left| \left( e^{j\psi}\mathcal{H} _{N_a}+\left| g \right| \right) xe^{j\angle g}-( e^{j\hat{\psi}}\mathcal{H} _{\hat{N}_a}+\left| g \right| ) \hat{x}e^{j\angle g} \right|^2 \nonumber\\
		&\quad \quad \quad -2\sqrt{P_t}\Re \left\{ w^{\ast}\left[ \left( e^{j\psi }\mathcal{H} _{N_a}+\left| g \right| \right) xe^{j\angle g}-( e^{j\hat{\psi}}\mathcal{H} _{\hat{N}_a}+\left| g \right| ) \hat{x}e^{j\angle g} \right] \right\} >0 \bigg)
	\end{align}
	\hrulefill
\end{figure*}
\setcounter{equation}{\value{TempEqCnt}}

\textit{Remark 1:} In RIS-CBC-APSK, when the A-Tx transmits unmodulated carrier waves, namely $A=1$ and $x\equiv 1$, the RIS can be considered as part of an access point, i.e., the combination of A-Tx and RIS can be regarded as an RIS-based transmitter with a $P$-APSK modulator.

\subsection{Performance Analysis}
In this subsection, we analyze the SER performance of passive RIS-CBC-APSK with ML detection. Rician fading is considered for all channel links, i.e., $h_i$, $f_{i}$, and $g_{i}$ are given by
\begin{align}\label{h_i}
	h_{i}=\sqrt{\rho_1}\left( \sqrt{\frac{K}{K+1}}+\sqrt{\frac{1}{K+1}}\tilde{h}_{i} \right), 
\end{align} 
\begin{align}\label{f_i}
	f_{i}=\sqrt{\rho_2}\left( \sqrt{\frac{K}{K+1}}+\sqrt{\frac{1}{K+1}}\tilde{f}_{i} \right), 
\end{align} 
and
\begin{align}\label{g_i}
	g=\sqrt{\rho_3}\left( \sqrt{\frac{K}{K+1}}+\sqrt{\frac{1}{K+1}}\tilde{g} \right), 
\end{align} 
where $\rho_1=\rho_r d_{1}^{-v_1}$, $\rho_2=\rho_r d_{2}^{-v_2}$, and $\rho_3=\rho_r d_{3}^{-v_3}$ denote the large-scale path loss of the A-Tx to RIS, RIS to C-Rx, and A-Tx to C-Rx links, respectively. Moreover, $\rho_r$ is the path-loss factor at the reference distance of $1$ meter (m), $v_1$ ($v_2$, $v_3$) represents the path-loss exponent of the A-Tx to RIS (RIS to C-Rx, A-Tx to C-Rx) link, and $d_{1}$ ($d_{2}$, $d_{3}$) denotes the distance between the A-Tx and the RIS (between the RIS and the C-Rx, between the A-Tx and the C-Rx). $K$ is the Rician factor. Finally, $\tilde{h}_{i}\sim \mathcal{C} \mathcal{N} \left( 0,1 \right)$, $\tilde{f}_{i}\sim \mathcal{C} \mathcal{N} \left( 0,1 \right)$, and $\tilde{g}\sim \mathcal{C} \mathcal{N} \left( 0,1 \right)$ denote the non-line-of-sight components of the A-Tx to RIS, RIS to C-Rx, and A-Tx to C-Rx links, respectively. 

From (\ref{ML_y}), the conditional pairwise error probability (PEP) can be written as (\ref{ConditionalPEP}), shown at the top of this page. In (\ref{ConditionalPEP}), the term on the left-hand side of ``$>$'' can be treated as a Gaussian random variable, denoted henceforth as $D$ with
\setcounter{equation}{15}
\begin{align}\label{E_D}
	\mathrm{E} \left\{ D \right\} &=-P_t\Big| \left( e^{j\psi}\mathcal{H} _{N_a}+\left| g \right| \right) x-( e^{j\hat{\psi}}\mathcal{H} _{\hat{N}_a}+\left| g \right| ) \hat{x} \Big|^2
\end{align} 
and
\begin{flalign}\label{Var_D}
	\mathrm{VAR} \left\{ D \right\} &=2P_tN_0\Big| \left( e^{j\psi}\mathcal{H} _{N_a}+\left| g \right| \right) x-( e^{j\hat{\psi}}\mathcal{H} _{\hat{N}_a}+\left| g \right| ) \hat{x} \Big|^2.
\end{flalign} 
With (\ref{E_D}) and (\ref{Var_D}), the conditional PEP can be expressed in the form of $Q$-function as
\begin{align}\label{PEP_Q_func}
	\mathrm{Pr}\left( \left( N_a, \psi,x \right) \rightarrow \left( \hat{N}_a,\hat{\psi},\hat{x} \right) |\left( \mathbf{h},\mathbf{f},g \right) \right) =Q\left( \sqrt{\frac{P_t\varLambda}{2N_0}} \right), 
\end{align} 
where $\varLambda=| e^{j\psi}\mathcal{H} _{N_a}x-e^{j\hat{\psi}}\mathcal{H} _{\hat{N}_a}\hat{x}+\left| g \right|( x-\hat{x} ) |^2$. 

To obtain the unconditional PEP, (\ref{PEP_Q_func}) should be further averaged over $\mathbf{h}$, $\mathbf{f}$, and $g$. Since $N_a\gg 1$, $\mathcal{H} _{N_a}$ can be approximated as a Gaussian random variable with the distribution $\mathcal{N} ( N_a\mu ,N_a\sigma ^2 ) $, where $\mu =\mathrm{E}\{\underline{h}_i\}=\mathrm{E}\{|h_{i}|\} \mathrm{E}\{|f_{i}|\}$, $\sigma^2=\mathrm{VAR}\{\underline{h}_i\}=\mathrm{E}\{|h_{i}|^2 \} \mathrm{E}\{|f_{i}|^2\} -\mathrm{E}\{|h_{i}| \} ^2\mathrm{E}\{| f_{i}|\} ^2 $,
\begin{align}\label{E_h_k}
	\mathrm{E}\left\{ \left| h_{i} \right|^k \right\} =&\left( 2b_{1} \right) ^{k/2}\exp \left( -{a_{1}^{2}}/({2b_{1}}) \right) \varGamma \left( 1+{k}/{2} \right)\nonumber\\  
	&\times\prescript{}{1}F_1\left( 1+{k}/{2};1;{a_{1}^{2}}/({2b_{1}}) \right), 
\end{align}
and
\begin{align}\label{E_f_k}
	\mathrm{E}\left\{ \left| f_{i} \right|^k \right\} =&\left( 2b_{2} \right) ^{k/2}\exp \left( -{a_{2}^{2}}/({2b_{2}}) \right) \varGamma \left( 1+{k}/{2} \right)\nonumber\\
	&\times \prescript{}{1}F_1\left( 1+{k}/{2};1;{a_{2}^{2}}/({2b_{2}}) \right), 
\end{align}
with $a_{1}=\sqrt{{\rho_{1}K}/({1+K})}$, $b_{1}={\rho_{1}}/({2+2K})$, $a_{2}=\sqrt{{\rho_{2}K}/({1+K})}$, $b_{2}={\rho_{2}}/({2+2K })$, and $k=1,2$. For large $K$, $|g|$ can be approximated as a Gaussian variable with the distribution $\mathcal{N}( \mu_g, \sigma_{g} ^2) $, where $\mu_g = \mathrm{E}\{|g|\}$, $\sigma _{g}^{2}=\mathrm{E}\{ | g |^2 \} -\mathrm{E}\{ | g | \} ^2$, and $ \mathrm{E}\{|g|^k\}$ can be similarly calculated as in (\ref{E_h_k}) or (\ref{E_f_k}).

\newcounter{TempEqCnt2} 
\setcounter{TempEqCnt2}{\value{equation}} 
\setcounter{equation}{23} 
\begin{figure*}[ht]	
	\vspace*{1pt}
	\begin{align}\label{sigma12_Na_smaller_Na_hat}
		\sigma_{12}^{2}=\sigma_{21}^{2}&=\frac{1}{2}\left\{\mathrm{VAR}\left\{\sum_{i=1}^{N_a}{2\underline{h}_i}+\sum_{i=\hat{N}_a-N_a+1}^{\hat{N}_a}{\underline{h}_i}\right\}-\mathrm{VAR}\left\{\sum_{i=1}^{N_a}{\underline{h}_i}\right\}-\mathrm{VAR}\left\{\sum_{i=1}^{\hat{N}_a}{\underline{h}_i}\right\}\right\}=N_a\sigma^2
	\end{align} 
\begin{align}\label{sigma12_Na_bigger_Na_hat}
	\sigma _{12}^{2}=\sigma _{21}^{2}&=\frac{1}{2}\left\{\mathrm{VAR}\left\{\sum_{i=1}^{\hat{N}_a}{2\underline{h}_i}+\sum_{i=N_a-\hat{N}_a+1}^{N_a}{\underline{h}_i}\right\}-\mathrm{VAR}\left\{\sum_{i=1}^{N_a}{\underline{h}_i}\right\}-\mathrm{VAR}\left\{\sum_{i=1}^{\hat{N}_a}{\underline{h}_i}\right\}\right\} =\hat{N}_a\sigma^2
\end{align}
	\hrulefill
\end{figure*}
\setcounter{equation}{\value{TempEqCnt2}}

By using an approximation to the $Q$-function \cite{Chiani_Q_func}, namely $Q( x ) \approx 1/12\cdot e^{-x^2/2} +1/4\cdot e^{-2x^2/3}$, the unconditional PEP can be written as
\begin{align}\label{PEP_ave}
	\Pr\left(\left(N_a,\psi,x\right)\rightarrow (\hat{N}_a,\hat{\psi},x)\right) 
	&\approx 1/12\cdot \mathcal{M}_{\varLambda} \left( -{P_t}/({4N_0}) \right)\nonumber\\ 
	& \hspace{-0.5cm}+ 1/4 \cdot \mathcal{M}_{\varLambda}\left( -{P_t}/({3N_0}) \right), 
\end{align}
where $\mathcal{M} _{\varLambda}\left( \cdot \right)$ is the moment generating function (MGF) of $\varLambda$. For calculating $\mathcal{M} _{\varLambda}\left( \cdot \right)$, we express $\varLambda$ in the quadratic form of Gaussian random variables, namely $\varLambda=\mathbf{a}^T\mathbf{B}\mathbf{a}$. For the case of $N_a \neq \hat{N}_a$, we have $\mathbf{a}=[\mathcal{H} _{N_a},\mathcal{H} _{\hat{N}_a},|g|]^T$ and $\mathbf{B}=\mathbf{b}_r{\mathbf{b}_r}^T+\mathbf{b}_i{\mathbf{b}_i}^T$, where $\mathbf{b}_r=[\Re\{xe^{j\psi}\},-\Re\{\hat{x}e^{j\hat{\psi}}\},\Re\{x-\hat{x}\}]^T$ and $\mathbf{b}_i=[\Im\{xe^{j\psi}\},-\Im\{\hat{x}e^{j\hat{\psi}}\},\Im\{x-\hat{x}\}]^T$. The mean vector and covariance matrix of $\mathbf{a}$ are computed as: 
\begin{align}\label{a_bar}
	\bar{\mathbf{a}}=\left[N_a\mu,\hat{N}_a\mu,\mu_g\right]^T
\end{align} 
and 
\begin{align}\label{A_tilde}
	\tilde{\mathbf{A}}=\left[ \begin{matrix}
		\sigma _{11}^{2}&		\sigma _{12}^{2}&		0\\
		\sigma _{21}^{2}&		\sigma _{22}^{2}&		0\\
		0&		0&		\sigma _{g}^{2}\\
	\end{matrix} \right],
\end{align} 
respectively, where $\sigma _{11}^{2}=N_a\sigma^2$ and $\sigma _{22}^{2}=\hat{N}_a\sigma^2$. The variances $\sigma _{12}^{2}$ and $\sigma_{21}^{2}$ are given by (\ref{sigma12_Na_smaller_Na_hat}) for $N_a<\hat{N}_a$ and (\ref{sigma12_Na_bigger_Na_hat}) for $N_a>\hat{N}_a$, where (\ref{sigma12_Na_smaller_Na_hat}) and (\ref{sigma12_Na_bigger_Na_hat}) are shown at the top of this page. Similarly, for the case of $N_a=\hat{N}_a$, we have $\mathbf{a}=[\mathcal{H} _{N_a},|g|]^T$, $\mathbf{b}_r=[\Re\{xe^{j\psi}-\hat{x}e^{j\hat{\psi}}\},\Re\{x-\hat{x}\}]^T$, and $\mathbf{b}_i=[\Im\{xe^{j\psi}-\hat{x}e^{j\hat{\psi}}\},\Im\{x-\hat{x}\}]^T$. Hence, the mean vector and covariance matrix of $\mathbf{a}$ are obtained as follows:
\setcounter{equation}{25}
\begin{align}\label{a_bar_equal}
	\bar{\mathbf{a}}=\left[N_a\mu,\mu_g\right]^T
\end{align} 
and 
\begin{align}\label{A_tilde_equal}
	\tilde{\mathbf{A}}=\left[ \begin{matrix}
		\sigma _{11}^{2}&		0\\
		0&			\sigma _{g}^{2}\\
	\end{matrix} \right],
\end{align} 
respectively, with $\sigma _{11}^{2}=N_a\sigma^2$.

After obtaining $\mathbf{B}$, $\tilde{\mathbf{A}}$, and $\mathbf{\bar{a}}$, we can derived the MGF of $\varLambda$ from \cite{Quadratic_form_RV} as
\begin{align}\label{MGF}
	\mathcal{M}_{\varLambda}\left( t \right) &=\left[ \det \left( \mathbf{I}-2t\mathbf{B}\tilde{\mathbf{A}} \right) \right] ^{-{1}/{2}}\nonumber\\
&\hspace{-0.4cm} \times\exp \left\{ -\frac{1}{2}\bar{\mathbf{a}}^T\left[ \mathbf{I}-\left( \mathbf{I}-2t\mathbf{B}\tilde{\mathbf{A}} \right) ^{-1} \right] \tilde{\mathbf{A}}^{-1}\bar{\mathbf{a}} \right\} . 
\end{align}

Finally, according to the union bounding technique \cite{SimonDigital}, the SER of active and backscatter information for passive RIS-CBC-APSK can be upper bounded by
\begin{align}\label{SER_A_PRIS}
	P_{\mathrm{SER}}^A\leq&\frac{1}{M}\sum_{N_a,\psi,x}{\sum_{\hat{N}_a,\hat{\psi},\hat{x}}{\Pr\left(\left(N_a,\psi,x\right)\rightarrow (\hat{N}_a,\hat{\psi},\hat{x})\right)}}\nonumber\\
	&\times\mathbb{I}(x\ne\hat{x}) 
\end{align}
and 
\begin{align}\label{SER_P_PRIS}
	P_{\mathrm{SER}}^P\leq&\frac{1}{M}\sum_{N_a,\psi,x}{\sum_{\hat{N}_a,\hat{\psi},\hat{x}}{\Pr\left(\left(N_a,\psi,x\right)\rightarrow (\hat{N}_a,\hat{\psi},\hat{x})\right)}}\nonumber\\
	&\times\mathbb{I}\left(\left(N_a,\psi\right)\ne(\hat{N}_a,\hat{\psi})\right),
\end{align}
respectively. Further, an upper bound on the SER of both active and backscatter information is given by 
\begin{align}\label{SER_total_PRIS}
	P_{\mathrm{SER}}\leq&\frac{1}{M}\sum_{N_a,\psi,x}{\sum_{\hat{N}_a,\hat{\psi},\hat{x}}{\Pr\left(\left(N_a,\psi,x\right)\rightarrow (\hat{N}_a,\hat{\psi},\hat{x})\right)}}\nonumber\\&\times\mathbb{I}\left(\left(N_a,\psi,x\right)\ne(\hat{N}_a,\hat{\psi},\hat{x})\right).
\end{align}

\textit{Remark 2:} From (\ref{PEP_Q_func}), we observe that the error performance of RIS-CBC-APSK is related to the Euclidean distances of the constellations of $x$ and $\mathcal{H} _{N_a}xe^{j\psi}$. When the direct link is blocked or weak, i.e., $g\cong0$, the error performance of RIS-CBC-APSK is dominated by the minimum Euclidean distance of the constituted APSK constellation, which is maximized in Section II.

\section{Proposed Active RIS-CBC with APSK}
In this section, we present an active RIS-CBC-APSK scheme. Unless otherwise specified, this section's mathematical definitions are the same as those in the previous Section~III.
\subsection{System Model}
The system model of active RIS-CBC-APSK is the same as that of passive RIS-CBC-APSK in Fig.~\ref{SystemModel_1}, except that the passive RIS is replaced with an active one. In this case, the received signal at the C-Rx can be expressed as 
\begin{align}\label{received_y_ARIS}
	y=\sqrt{P_t}\left( \mathbf{f}^T\mathbf{\Phi}\mathbf{h}+g \right) x+\mathbf{f}^T\mathbf{\Phi}\mathbf{v}+w,
\end{align}
where $\mathbf{v}=[ v_1,\ldots ,v_N ] ^T$ with $v_i\sim\mathcal{C} \mathcal{N} \left( 0,N_v \right)$ is the thermal noise generated by the power amplifier at each $i$-th RIS element.

\newcounter{TempEqCnt3} 
\setcounter{TempEqCnt3}{\value{equation}} 
\setcounter{equation}{40} 
\begin{figure*}[ht]	
	\vspace*{1pt}
	\begin{align}\label{sigma12_Na_smaller_Na_hat_activeRIS}
		\sigma _{12}^{2}=\sigma _{21}^{2}&=\frac{1}{2}\left\{\mathrm{VAR}\left\{\sum_{i=1}^{N_a}{2\xi\underline{h}_i}+\sum_{i=N_a+1}^{\hat{N}_a}{(\xi+1)\underline{h}_i}+\sum_{i=\hat{N}_a+1}^{N}{2\underline{h}_i}\right\}-\mathrm{VAR}\left\{\sum_{i=1}^{N_a}{\xi\underline{h}_i}+\sum_{i=N_a+1}^{N}{\underline{h}_i}\right\} \nonumber \right.\\ 
		&\quad \left.-\mathrm{VAR}\left\{\sum_{i=1}^{\hat{N}_a}{\xi\underline{h}_i}+\sum_{i=\hat{N}_a+1}^{N}{\underline{h}_i}\right\}\right\}=\left[N_a\xi^2+\left(\hat{N}_a-N_a\right)\xi+N-\hat{N}_a\right]\sigma^2
	\end{align} 
\begin{align}\label{sigma12_Na_bigger_Na_hat_activeRIS}
	\sigma _{12}^{2}=\sigma _{21}^{2}&=\frac{1}{2}\left\{\mathrm{VAR}\left\{\sum_{i=1}^{\hat{N}_a}{2\xi\underline{h}_i}+\sum_{i=\hat{N}_a+1}^{N_a}{(\xi+1)\underline{h}_i}+\sum_{i=N_a+1}^{N}{2\underline{h}_i}\right\}-\mathrm{VAR}\left\{\sum_{i=1}^{N_a}{\xi\underline{h}_i}+\sum_{i=N_a+1}^{N}{\underline{h}_i}\right\}\nonumber \right.\\ 
	&\quad \left.-\mathrm{VAR}\left\{\sum_{i=1}^{\hat{N}_a}{\xi\underline{h}_i}+\sum_{i=\hat{N}_a+1}^{N}{\underline{h}_i}\right\}\right\}=\left[\hat{N}_a\xi^2+\left(N_a-\hat{N}_a\right)\xi+N-N_a\right]\sigma^2
\end{align}
	\hrulefill
\end{figure*}
\setcounter{equation}{\value{TempEqCnt3}}

Particularly, in active RIS-CBC-APSK, each RIS element can be configured as active or passive by turning ON or OFF the power amplifier, i.e., $\beta_i \in \{1,\xi\}$ with $\xi>1$ for $i=1,\ldots,N$. The amplitude coefficient is $\xi$ for each active element and one for each passive element. Let $N_a$ and $N_p$ with $N_a+N_p=N$ denote the number of active and passive elements, respectively. As in passive RIS-CBC-APSK, $\phi_i$ is set as (\ref{thata_value}). Then, (\ref{received_y_ARIS}) can be rewritten as follows:
\begin{align}\label{receieved_y2_ARIS}
	y=&\sqrt{P_t}\left( \xi\mathcal{H} _{N_a}e^{j\psi}+\mathcal{H} _{N_p}e^{j\psi}+\left| g \right| \right) xe^{j\angle g}+w_e, 
\end{align}
where $\mathcal{H} _{N_p}=\sum\nolimits_{i=N_a+1}^{N}{\underline{h}_i}$, and $w_e=\xi e^{j\psi}\sum_{i=1}^{N_a}{|f_i|e^{j(-\angle h_i+\angle g)}v_i}+w$ is the overall noise term. Since $N_a\gg 1$ in practice, $w_e$ can be approximated as a Gaussian random variable with the distribution $\mathcal{C} \mathcal{N} ( 0,N_e )$, where $N_e=\xi ^2N_a\mathrm{E}\{ |f_i|^2\} N_v+N_0$. 

Similar to passive RIS-CBC-APSK, active RIS-CBC-APSK embeds backscatter information into $(N_a,\psi)$ and makes $(\xi\mathcal{H} _{N_a}+\mathcal{H} _{N_p})xe^{j\psi}$ be an APSK symbol. By considering that the active and passive elements reflect incident signals with amplitude coefficients $\xi$ and $1$, the set of all possible values of $N_a$ is given by 
\begin{align}\label{Na_ARIS}
	\mathcal{N}_a&=\Bigg\{\lfloor \frac{\xi N/\prod\nolimits_{\varsigma =1}^{R-1}{\gamma _{\varsigma}}-N}{\xi -1} \rceil,\lfloor \frac{\xi N/\prod\nolimits_{\varsigma =2}^{R-1}{\gamma _{\varsigma}}-N}{\xi -1} \rceil, \nonumber \\
	& \quad \quad \ldots,N\Bigg\},
\end{align}
where $\xi\geq \prod_{\varsigma=1}^{R-1}{\gamma _\varsigma}$.

At the C-Rx, the active and backscatter information can be recovered by estimating $x$ and $(N_a, \psi)$ jointly with ML detection, i.e.,
\begin{align}\label{ML_y_ARIS}
	&\left( \hat{N}_a ,\hat{\psi},\hat{x} \right) \nonumber\\&= \mathrm{arg} \min_{ N_a ,\psi,x } \Big| y-\sqrt{P_t}\left( \xi\mathcal{H} _{N_a}e^{j\psi}+\mathcal{H} _{N_p}e^{j\psi}+\left| g \right| \right) xe^{j\angle g} \Big|^2, 
\end{align}

\textit{Remark 3:} As seen from (\ref{Na_ARIS}), the amplitude coefficient should satisfy $\xi\geq \prod_{\varsigma=1}^{R-1}{\gamma _\varsigma}$ to ensure that each entity in $\mathcal{N}_a$ is greater than 0. When the constraint cannot be met, the passive elements can be turned OFF, and thus, they will not reflect incident signals. In this case, $\mathcal{N}_a$ is the same as (\ref{N_a}). We note that these two modes of active RIS-CBC-APSK share the same resulting APSK constellation.

\subsection{Performance Analysis}
According to (\ref{ML_y_ARIS}), the conditional PEP is given by
\begin{align}\label{ConditionalPEP_ARIS}
	&\mathrm{Pr}\left( \left( N_a, \psi,x \right) \rightarrow \left( \hat{N}_a,\hat{\psi},\hat{x} \right) |\left( \mathbf{h},\mathbf{f},g \right) \right)\nonumber\\ &=\mathrm{Pr}\bigg( \left| y-\sqrt{P_t}\left( e^{j\psi}\xi\mathcal{H} _{N_a}+e^{j\psi}\mathcal{H} _{N_p}+\left| g \right| \right) xe^{j\angle g} \right|^2 \nonumber \\
	&\quad >\left| y-\sqrt{P_t}\left( e^{j\hat{\psi}}\xi\mathcal{H} _{\hat{N}_a}+e^{j\hat{\psi}}\mathcal{H} _{\hat{N}_p}+\left| g \right| \right) \hat{x}e^{j\angle g} \right|^2 \bigg) \nonumber \\
	&= Q\left(\sqrt{\frac{P_t\varXi}{2N_e}}\right),
\end{align}
where $\varXi=\big| (\xi\mathcal{H} _{N_a}+\mathcal{H} _{N_p})xe^{j\psi}-(\xi\mathcal{H} _{\hat{N}_a}+\mathcal{H} _{\hat{N}_p})\hat{x}e^{j\hat{\psi}} +\left| g \right|\left( x-\hat{x} \right) \big|^2$.

By averaging (\ref{ConditionalPEP_ARIS}) over $\mathbf{h}$, $\mathbf{f}$, and $g$, the unconditional PEP can be approximated as 
\begin{align}\label{PEP_ave_ARIS}
	\Pr\left(\left(N_a,\psi,x\right)\rightarrow (\hat{N}_a,\hat{\psi},x)\right) &\approx 1/12\cdot \mathcal{M}_{\varXi} \left( -{P_t}/({4N_e}) \right) \nonumber\\
	& \hspace{-0.5cm}+ 1/4 \cdot \mathcal{M}_{\varXi}\left( -{P_t}/({3N_e}) \right), 
\end{align}
where $\mathcal{M} _{\varXi}\left( \cdot \right)$ is the MGF of $\varXi$. Also, for calculating $\mathcal{M} _{\varXi}\left( \cdot \right)$, we express $\varXi$ in the quadratic form of Gaussian random variables, namely $\varXi=\mathbf{c}^T\mathbf{B}\mathbf{c}$. For the case of $N_a \neq \hat{N}_a$, we have $\mathbf{c}=[\xi\mathcal{H} _{N_a}+\mathcal{H} _{N_p},\xi\mathcal{H}_{\hat{N}_a}+\mathcal{H} _{\hat{N}_p},|g|]^T$ and $\mathbf{B}=\mathbf{b}_r{\mathbf{b}_r}^T+\mathbf{b}_i{\mathbf{b}_i}^T$ with $\mathbf{b}_r=[\Re\{xe^{j\psi}\},-\Re\{\hat{x}e^{j\hat{\psi}}\},\Re\{x-\hat{x}\}]^T$ and $\mathbf{b}_i=[\Im\{xe^{j\psi}\},-\Im\{\hat{x}e^{j\hat{\psi}}\},\Im\{x-\hat{x}\}]^T$. The mean vector and covariance matrix of $\mathbf{c}$ are computed as follows:
\begin{align}\label{c_bar}
	\bar{\mathbf{c}}=\left[\left[(\xi-1)N_a+N\right]\mu,\left[(\xi-1)\hat{N}_a+N\right]\mu,\mu_g\right]^T
\end{align} 
and 
\begin{align}\label{C_tilde}
	\mathbf{\tilde{C}}=\left[ \begin{matrix}
		\sigma _{11}^{2}&		\sigma _{12}^{2}&		0\\
		\sigma _{21}^{2}&		\sigma _{22}^{2}&		0\\
		0&		0&		\sigma _{g}^{2}\\
	\end{matrix} \right],
\end{align} 
respectively, where $\sigma _{11}^{2}=[(\xi-1)N_a+N]\sigma^2$ and $\sigma _{22}^{2}=[(\xi-1)\hat{N}_a+N]\sigma^2$. The variances $\sigma _{12}^{2}$ and $\sigma _{21}^{2}$ are given by (\ref{sigma12_Na_smaller_Na_hat_activeRIS}) for $N_a<\hat{N}_a$ and (\ref{sigma12_Na_bigger_Na_hat_activeRIS}) for $N_a>\hat{N}_a$, where (\ref{sigma12_Na_smaller_Na_hat_activeRIS}) and (\ref{sigma12_Na_bigger_Na_hat_activeRIS}) are shown at the top of this page. Similarly, for the case of $N_a=\hat{N}_a$, we have $\mathbf{c}=[\xi\mathcal{H} _{N_a}+\mathcal{H} _{N_p},|g|]^T$, $\mathbf{b}_r=[\Re\{xe^{j\psi}-\hat{x}e^{j\hat{\psi}}\},\Re\{x-\hat{x}\}]^T$, and $\mathbf{b}_i=[\Im\{xe^{j\psi}-\hat{x}e^{j\hat{\psi}}\},\Im\{x-\hat{x}\}]^T$. Hence, the mean vector and covariance matrix of $\mathbf{c}$ are given by
\setcounter{equation}{42} 
\begin{align}\label{c_bar_equal}
	\bar{\mathbf{c}}=\left[\left[(\xi-1)N_a+N\right]\mu,\mu_g\right]^T
\end{align} 
and 
\begin{align}\label{C_tilde_equal}
	\tilde{\mathbf{C}}=\left[ \begin{matrix}
		\sigma _{11}^{2}&		0\\
		0&			\sigma _{g}^{2}\\
	\end{matrix} \right],
\end{align} 
respectively, with $\sigma _{11}^{2}=[(\xi-1)N_a+N]\sigma^2$.

After obtaining $\mathbf{B}$, $\tilde{\mathbf{C}}$, and $\bar{\mathbf{c}}$, we can obtain the MGF of $\varLambda$ like (\ref{MGF}). Finally, the upper bounds on the SERs of detecting $x$ and $(N_a,\psi)$ can be derived by putting (\ref{PEP_ave_ARIS}) into (\ref{SER_A_PRIS}) and (\ref{SER_P_PRIS}), respectively.

\section{LC detector and Extension to MISO Systems}
In this section, we investigate the LC detector and beamforming designs for RIS-CBC-APSK and its MISO systems, respectively.
\subsection{LC Detector}
\begin{figure}[t]
	\centering
	\includegraphics[width=2.5in]{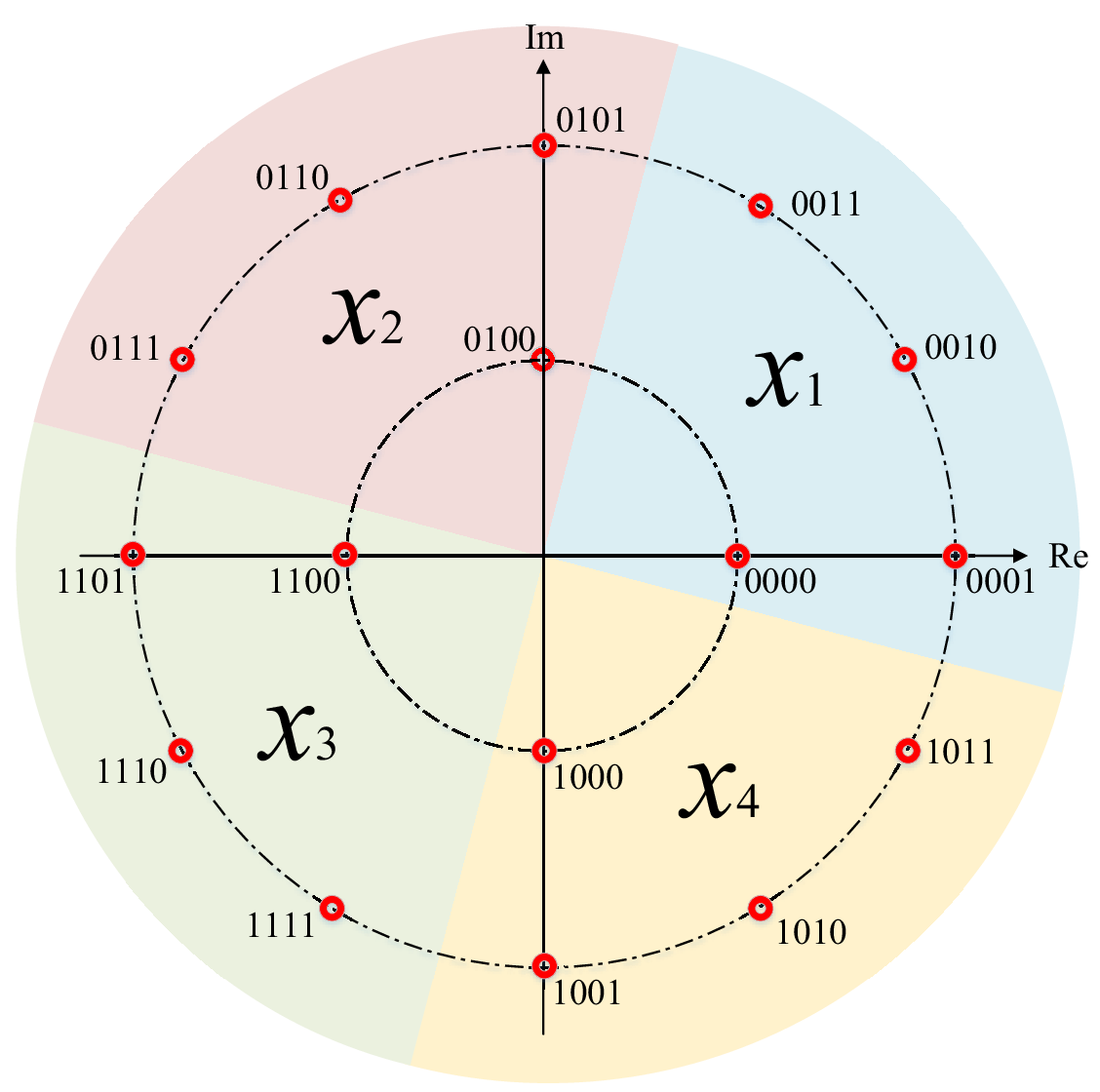}
	\caption{The partition for $4+12$-APSK with $A=P=4$.}
	\label{Const_partition}
\end{figure}
In this subsection, we propose a LC detector to separately retrieve $x$ and $(N_a,\psi)$ from $y$. Only the LC detector for passive RIS-CBC-APSK is illustrated here, since the detection process for the active counterpart is similar. 

By recalling the constitution of RIS-CBC-APSK constellations, it is easy to conclude that every constellation can be partitioned into $A$ regions uniformly according to the active information. An example of the partition for $4+12$-APSK with $A=4$ and $P=4$ is presented in Fig.~\ref{Const_partition}, where $x_n$ is the $n$-th constellation point of $\mathcal{A}$ with $n=1,2,\ldots,A $. Therefore, active information can be detected first by determining which region the received signal is located in. To begin with, we define an anchor vector 
\begin{align}\label{E_passive_info_1}
	\bar{\mathbf{q} } &= \mathrm{E}\left\{ \boldsymbol{\beta}e^{j\psi} \right\} \nonumber \\
	&=\frac{1}{P}\left( \boldsymbol{\beta }_1e^{j\psi _1}+\boldsymbol{\beta }_2e^{j\psi _2}+\cdots +\boldsymbol{\beta }_Pe^{j\psi _P} \right),
\end{align}
where $\boldsymbol{\beta}=[\beta_1,\ldots,\beta_N]^T$ and $\boldsymbol{\beta }_ne^{j\psi _n}$ represents the $n$-th backscatter information symbol with $n= 1,2,\ldots ,P$. Then, by means of the anchor vector, the active information can be recovered from
\begin{align}\label{LC_active_Info}
	\hat{x} = \mathrm{arg} \min_{ x } \left| y-\sqrt{P_t}\left( \bar{\mathbf{q} }^T\mathbf{\Theta \bar{h}}  + g \right) x \right|^2,
\end{align}
where $\mathbf{\bar{h}}=\mathrm{diag}\{ \mathbf{f}\}\mathbf{h}$ and $\mathbf{\Theta}=\mathrm{diag}\{\boldsymbol{\theta}\}$ with $\boldsymbol{\theta}=[ \theta _1,\ldots ,\theta _N]^T = [e^{j(\phi_1-\psi)},\ldots ,e^{j(\phi_N-\psi)}]^T$. However, since we use $\bar{\mathbf{q} }$ rather than the exact backscatter information in (\ref{LC_active_Info}), the detection performance is not satisfactory. To solve this problem, we revise (\ref{LC_active_Info}) as 
\begin{align}\label{LC_active_Info_sort}
	&\left(\hat{x}_1,\ldots,\hat{x}_I,\ldots,\hat{x}_A \right) \nonumber\\&= \underset{x}{\mathrm{arg\ sort}} \left(\left| y-\sqrt{P_t}\left( \mathbf{q}^T\mathbf{\Theta \bar{h}}  +  g  \right) x \right|^2,\uparrow\right),
\end{align}
where $\hat{x}_I$ is the $I$-th element after the operation in (\ref{LC_active_Info_sort}) with $1\leq I <A$. The set of detection candidates can be obtained as $\chi^{*}=\{\hat{x}_1,\hat{x}_2,\ldots,\hat{x}_I\}$. Afterwards, with $\chi^{*}$, the joint detection of $x$ and $(N_a,\psi)$ can be formulated as follows:
\begin{align}\label{LC_Joint}
	\left(\hat{x},\hat{N}_a,\hat{\psi}\right) = \mathrm{arg} \min_{ x\in \chi^{*},N_a,\psi } \left| y-\sqrt{P_t}\left( e^{j\psi}\boldsymbol{\beta }^T\mathbf{\Theta \bar{h}}  + g \right) x \right|^2.
\end{align}

We observe that the computational complexity of LC detection is of order $~\mathcal{O}(IP+A)$. When $I\leq A/2$, the detection complexity of LC detection is always lower than that of ML detection. It can be observed from Section VI that, when $I$ increases to $A/2$, the performance of LC detection is almost the same with that of ML detection. Therefore, the proposed LC detection is able to achieve different trade-offs between detection performance and complexity by adjusting the value of $I$.

\subsection{Extension to MISO Systems}
Both passive and active RIS-CBC-APSK can be extended to MISO systems, where the A-Tx and the C-Rx are equipped with $N_t$ antennas and a single antenna, respectively. Since the extensions for passive and active RIS-CBC-APSK are similar, we take passive RIS-CBC-APSK as an illustrative example without loss of generality. 

For passive RIS-CBC-APSK-MISO, the received signal at the C-Rx can be written as 
\begin{align}\label{y_MISO}
	y=\sqrt{P_t}\left( \mathbf{f}^T\mathbf{\Phi H}+\mathbf{g}^T \right)\mathbf{p} x+w,
\end{align}
where $\mathbf{p}\in \mathbb{C} ^{N_t\times 1}$ denotes the active beamforming at the A-Tx with $\lVert \mathbf{p} \rVert ^2\leq 1$, $\mathbf{g}\in \mathbb{C}^{N_t\times 1}$ is the channel vector between the A-Tx and C-Rx, and $\mathbf{H}\in \mathbb{C} ^{N\times N_t}$ is the channel matrix between the A-Tx and RIS. With $\boldsymbol{\beta}$ and $\mathbf{\Theta}$, (\ref{y_MISO}) can be rewritten as
\begin{align}\label{y_MISO2}
	y=\sqrt{P_t}\left( e^{j\psi}\boldsymbol{\beta }^T\mathbf{\Theta \bar{H}}+\mathbf{g}^T \right) \mathbf{p}x+w,
\end{align}
where $\mathbf{\bar{H}}=\mathrm{diag}\{ \mathbf{f}\}\mathbf{H}$. The active and passive beamforming can be jointly optimized to maximize the average received power, which is given by
\begin{align}\label{Ave_received_signal_power}
	&\mathrm{E}\left\{ \left| \left( e^{j\psi}\boldsymbol{\beta }^T\mathbf{\Theta \bar{H}}+\mathbf{g}^T \right) \mathbf{p} \right|^2 \right\}  \nonumber\\
	&=\mathbf{p}^H \Big( \mathbf{\bar{H}}^H\mathbf{\Theta }^H\mathrm{E}\left\{ \boldsymbol{\beta \beta }^T \right\} \mathbf{\Theta \bar{H}}+\mathbf{\bar{H}}^H\mathbf{\Theta }^H\mathrm{E}\left\{ \boldsymbol{\beta}e^{-j\psi} \right\} \mathbf{g}^T \nonumber\\
	&\quad +\mathbf{g}^*\mathrm{E}\left\{ \boldsymbol{\beta}^T e^{j\psi}\right\}\mathbf{\Theta \bar{H}}+\mathbf{g}^*\mathbf{g}^T\Big)\mathbf{p}
\end{align}
with $\mathbf{\bar{H}}=\mathrm{diag}\{ \mathbf{f}\}\mathbf{H}$ and $\boldsymbol{\beta}=[\beta_1,\ldots,\beta_N]^T$. Note that the expectations in (\ref{Ave_received_signal_power}) are with respect to the transmitted backscatter information that is embedded in $\boldsymbol{\beta}$ and $\psi$. Let $\mathbf{\bar{u}}=\mathrm{E}\left\{ \boldsymbol{\beta}e^{-j\psi} \right\}$ and $\mathbf{U}=\mathrm{E}\left\{ \boldsymbol{\beta \beta }^T \right\}$, i.e.,
\begin{align}\label{E_passive_info}
	\mathbf{\bar{u}}=\frac{1}{P}\left( \boldsymbol{\beta }_1e^{-j\psi _1}+\boldsymbol{\beta }_2e^{-j\psi _2}+\cdots +\boldsymbol{\beta }_Pe^{-j\psi _P} \right)
\end{align}
and 
\begin{align}\label{E_passive_info_mat}
	\mathbf{U}=\frac{1}{P}\left( \boldsymbol{\beta }_1\boldsymbol{\beta }_1^T+\boldsymbol{\beta }_2\boldsymbol{\beta }_2^T+\cdots +\boldsymbol{\beta }_P\boldsymbol{\beta }_P^T \right).
\end{align}
The optimization problem can be thus formulated as
\begin{align}
	\text{(P3)}: 	\underset{ \mathbf{p},\mathbf{\Theta }}{\text{max}}\ &~ \mathbf{p}^H\left[ \begin{array}{c}
		\mathbf{\Theta \bar{H}}\\
		\mathbf{g}^T\\
	\end{array} \right] ^H\left[ \begin{matrix}
		\mathbf{U}&		\mathbf{\bar{u}}\\
		\mathbf{\bar{u}}^H&		1\\
	\end{matrix} \right] \left[ \begin{array}{c}
		\mathbf{\Theta \bar{H}}\\
		\mathbf{g}^T\\
	\end{array} \right] \mathbf{p}\label{P3} \\
	{s.t.} \quad&~
	\lVert \mathbf{p} \rVert ^2\leq 1\tag{\ref{P3}{a}} \label{P3a}\\ 
	&~
	|\theta_i|=1,\quad i=1,\ldots ,N\tag{\ref{P3}{b}} \label{P3b}
\end{align}
Obviously, problem (P3) is a non-convex problem. In what follows, the active and passive beamforming are optimized alternatively.

\subsubsection{Active Beamforming Design}

For any given $\mathbf{\Theta}$, problem (P3) can be rewritten as 
\begin{align}
	\text{(P3.1)}: 	\underset{ \mathbf{p}}{\text{max}}\ &~ \mathbf{p}^H\mathbf{V} \mathbf{p}\label{P3.1} \\
	s.t. \quad&~
	\lVert \mathbf{p} \rVert ^2\leq 1\tag{\ref{P3.1}{a}} \label{P3.1a},
\end{align}
where
\begin{align}
	\mathbf{V}=\left[ \begin{array}{c}
		\mathbf{\Theta \bar{H}}\\
		\mathbf{g}^T\\
	\end{array} \right] ^H\left[ \begin{matrix}
		\mathbf{U}&		\mathbf{\bar{u}}\\
		\mathbf{\bar{u}}^H&		1\\
	\end{matrix} \right] \left[ \begin{array}{c}
		\mathbf{\Theta \bar{H}}\\
		\mathbf{g}^T\\
	\end{array} \right].
\end{align}
It is easy to obtain the optimal solution to problem (P3.1), i.e., $\mathbf{p}=\mathbf{e}_v/\lVert \mathbf{e}_v \rVert$, where $\mathbf{e}_v$ denotes the eigenvector corresponding to the maximum eigenvalue of $\mathbf{V}$.

\subsubsection{Passive Beamforming Design}
For any given $\mathbf{p}$, by omitting some irrelevant terms, problem (P3) can be rewritten as 
\begin{align}
	\text{(P3.2)}: 	\underset{ \boldsymbol{\theta}}{\text{max}}\ &~ \boldsymbol{\theta }^H\mathbf{G}^H\mathbf{UG}\boldsymbol{\theta }+\tilde{g}\boldsymbol{\theta }^H\mathbf{G}^H\mathbf{\bar{u}}+\tilde{g}^*\mathbf{\bar{u}}^H\mathbf{G}\boldsymbol{\theta }
	\label{P3.2} \\
	{s.t.} \quad&~
	|\theta_i|=1, \quad i=1,\ldots ,N,\tag{\ref{P3.2}{a}} \label{P3.2a}
\end{align}
where $\mathbf{G}=\mathrm{diag}\{\mathbf{\bar{H}p}\}$ and $\tilde{g}=\mathbf{g}^T\mathbf{p}$. Problem (P3.2) is a non-convex quadratically constrained quadratic program (QCQP). To solve it, we reformulate (P3.2) as a homogeneous QCQP by introducing an auxiliary variable $t$ with $|t|=1$, i.e., 
\begin{align}
	\text{(P3.3)}: 	\underset{ \boldsymbol{\tilde{\theta}}}{\text{max}}\ &~ \boldsymbol{\tilde{\theta}}^H\mathbf{R} \boldsymbol{\tilde{\theta}}
	\label{P3.3} \\
	{s.t.} \quad&~
	|\tilde{\theta}_i|=1, \quad i=1,\ldots ,N\tag{\ref{P3.3}{a}} \label{P3.3a}.
\end{align}
where
\begin{align}
	\mathbf{R}=\left[ \begin{matrix}
		\mathbf{G}^H\mathbf{UG}&		\tilde{g}\mathbf{G}^H\mathbf{\bar{u}}\\
		\tilde{g}^*\mathbf{\bar{u}}^H\mathbf{G}&		0\\
	\end{matrix} \right], \quad \boldsymbol{\tilde{\theta}}=\left[ \begin{array}{c}
		\boldsymbol{\theta }\\
		t\\
	\end{array} \right]  .
\end{align}
We observe that $\mathbf{\tilde{\Theta}}=\boldsymbol{\tilde{\theta}}\boldsymbol{\tilde{\theta}}^H$ is positive semi-define with $\mathrm{rank}(\mathbf{\tilde{\Theta}})=1$. Hence, by applying the semi-definite relaxation method \cite{Z_SDR}, problem (P3.3) simplifies to
\begin{align}
	\text{(P3.4)}: 	\underset{ \mathbf{\tilde{\Theta}}}{\text{max}}\ &~ \mathrm{trace}(\mathbf{R}\mathbf{\tilde{\Theta}})
	\label{P3.4} \\
	{s.t.} \quad&~
	\tilde{\Theta}_{ii}=1, \quad i=1,\ldots ,N+1\tag{\ref{P3.4}{a}} \label{P3.4a},\\
	\quad&~
	\mathbf{\tilde{\Theta}} \succeq 0\tag{\ref{P3.4}{b}} \label{P3.4b}.
\end{align}
It can be seen that problem (P3.4) is a convex semi-definite program that can be solved by existing convex optimization solvers, such as CVX. However, the derived $\mathbf{\tilde{\Theta}}$ may not satisfy $\mathrm{rank}(\mathbf{\tilde{\Theta}})=1$. To solve this problem, we let $\boldsymbol{\tilde{\theta}}^{'}=[ \tilde{\theta}_{1}^{'},\ldots ,\tilde{\theta}_{N+1}^{'} ] ^T=\sqrt{\lambda _{\tilde{\theta}}}\boldsymbol{e}_{\tilde{\theta}}$ \cite{Z_SDR} and take $\boldsymbol{\theta }^{\dag}=[ \exp(j\angle(\tilde{\theta}_{1}^{'}/\tilde{\theta}_{N+1}^{'})),\ldots ,\exp(j\angle(\tilde{\theta}_{N}^{'}/\tilde{\theta}_{N+1}^{'})) ] ^T$ as a sub-optimal solution to problem (P3.2), where $\lambda _{\tilde{\theta}}$ is the maximum eigenvalue of $\mathbf{\tilde{\Theta}}$ and $\boldsymbol{e}_{\tilde{\theta}}$ is the corresponding eigenvector.

The alternating optimization (AO) algorithm proceeds by alternately solving problems (P3.1) and (P3.4) until the convergence criterion is met or the maximum number of iterations is reached. The complete algorithm is summarized in Algorithm 1, where  $\epsilon$ and $\mathcal{K}$ denote the convergence criterion and the maximum number of iterations, respectively.

\begin{algorithm}[!t]
	\caption{Active and Passive Beamforming for RIS-CBC-APSK}
	\label{Algorithm 1}
	\begin{algorithmic}[1]
		\STATE Input: $\mathbf{H}$, $\mathbf{f}$, $\mathbf{g}$, $\epsilon$, and $\mathcal{K}$.
		\STATE Initialization: $\boldsymbol{\theta }^{(0)}=\mathbf{1}_{N\times 1}$ and $k=1$.
		\REPEAT
		\STATE $\mathbf{p}^{(k)}=\mathbf{e}^{(k)}_v/\lVert \mathbf{e}^{(k)}_v \rVert$;\\
		\STATE $\boldsymbol{\theta }^{\dag}=\left[ \exp(j\angle(\tilde{\theta}_{1}^{'}/\tilde{\theta}_{N+1}^{'})),\ldots ,\exp(j\angle(\tilde{\theta}_{N}^{'}/\tilde{\theta}_{N+1}^{'})) \right] ^T$;\\
		\STATE Update $k=k+1$.\\
		\UNTIL $\left|| \mathbf{p}^{(k)H}\mathbf{V}^{(k)}\mathbf{p}^{(k)}|^2-| \mathbf{p}^{(k-1)H}\mathbf{V}^{(k-1)}\mathbf{p}^{(k-1)} |^2 \right|<\epsilon$ or $k=\mathcal{K}$.
		\STATE Output: $\mathbf{p}^{(k)}$ and $\boldsymbol{\theta}^{(k)}$.
	\end{algorithmic}
\end{algorithm}

\begin{figure*}[t]
	\centering
	\includegraphics[width=7in]{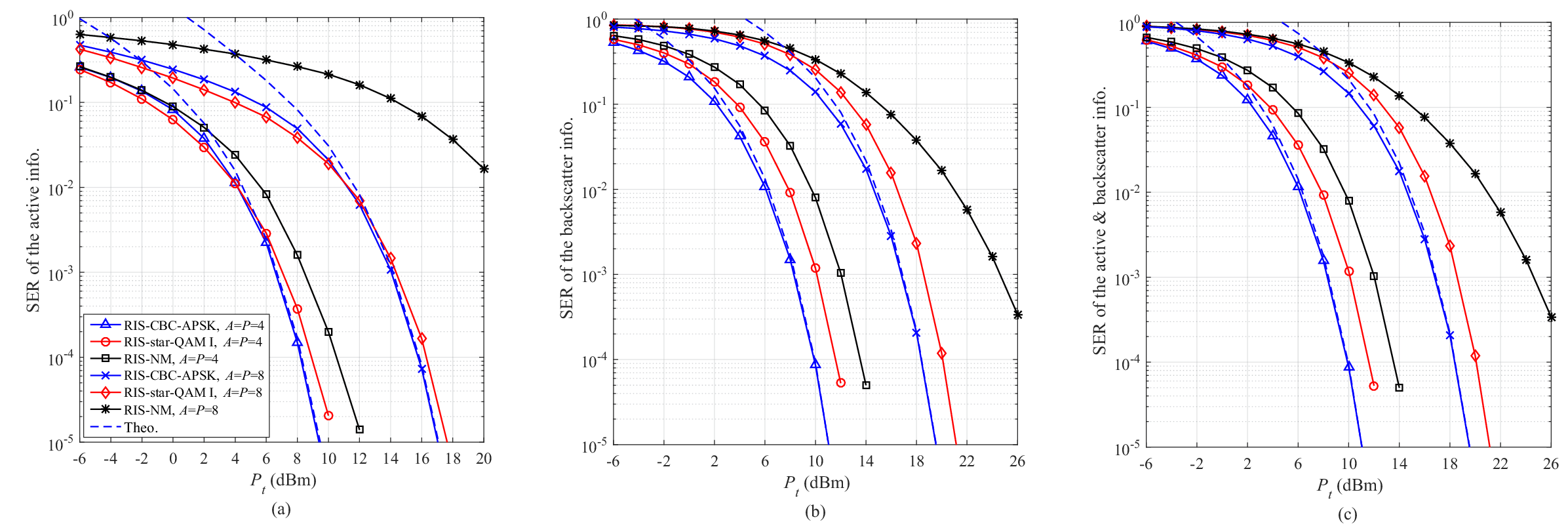}
	\caption{Performance comparison among passive RIS-CBC-APSK, RIS-star-QAM I, and RIS-NM with $A=P=4$ (4 bpcu) and $A=P=8$ (6 bpcu): (a) SER of the active information, (b) SER of the backscatter information, and (c) SER of the active and backscatter information.}
	\label{passive_RIS_active_passive_comparison}
\end{figure*}

\section{Simulation Results and Discussion}
In this section, we conduct Monte Carlo simulations to evaluate the SER performance of the proposed RIS-CBC-APSK schemes. In all simulations, $\gamma_1,\ldots,\gamma_{R-1}$ are obtained via computer search with the step size of $0.01$ and the SER is depicted as a function of $P_t$. The noise power is set as $N_0=N_v=-80$ dBm and the total number of RIS elements is set to $N=128$. RIS-CBC-APSK employs $4+12$-APSK, $4+12+16$-APSK, and $8+24+32$-APSK for $M=16,32$, and $64$, respectively. The other parameters are set as follows: $d_1=5$ m, $d_2=50$ m, $d_3=54$ m, $K=8$, $v_1=2.0$, $v_2=2.2$, $v_3=3.5$,  $\rho_r=-30$ dBm, $\epsilon=10^{-10}$, and $\mathcal{K}=4$. All schemes use the optimal ML detection unless otherwise specified. All simulation results are obtained by averaging over at least $10^5$ channel realizations. 

\begin{figure}[t]
	\centering
	\includegraphics[width=3.5in]{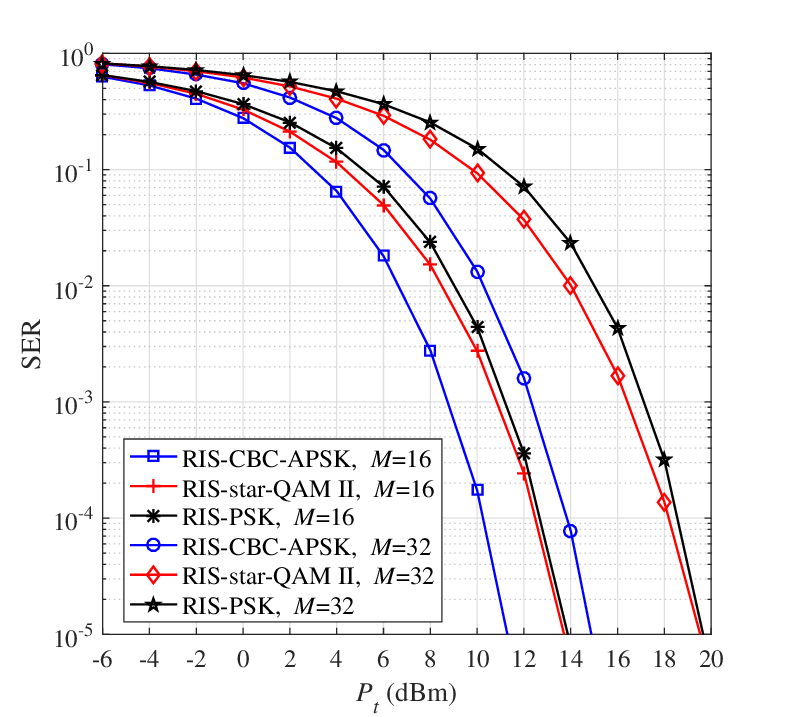}
	\caption{Performance comparison among passive RIS-CBC-APSK, RIS-star-QAM II, and RIS-PSK schemes with $A\equiv 1$ and $P=M=16,32$, where the direct A-Tx to C-Rx link is ignored.}
	\label{RIS_CBC_APSK_transmitter}
\end{figure}

Figure~\ref{passive_RIS_active_passive_comparison} depicts the SER performance comparison among passive RIS-CBC-APSK, RIS-star-QAM I\footnote{In both \cite{Wu_RIS_star_QAM} and \cite{Li_RIS_APSK}, star-QAM constellations are designed for RIS-CBC. For ease of distinction, the schemes in \cite{Wu_RIS_star_QAM} and \cite{Li_RIS_APSK} are named as RIS-star-QAM I and RIS-star-QAM II, respectively. Note that RIS-star-QAM II in \cite{Li_RIS_APSK} only applies to the case where the A-Tx transmits unmodulated carrier waves, which will be evaluated in Fig.~\ref{RIS_CBC_APSK_transmitter}.} \cite{Wu_RIS_star_QAM}, and RIS-NM \cite{Li_NM} schemes at the data rates of 4 bpcu ($A=P=4$) and 6 bpcu ($A=P=8$). As seen from Figs.~\ref{passive_RIS_active_passive_comparison}(a)-(c), the theoretical curves agree with the simulated counterparts very well, verifying the analysis presented in Section III.B. As expected, increasing $A$ and $P$ results in the worse SER performance for both active and backscatter transmission of all considered schemes. For both data rates of 4 bpcu and 6 bpcu, passive RIS-CBC-APSK performs better than RIS-NM in terms of both active and backscatter transmission throughout the whole $P_t$ region. Moreover, the performance improvement for 6 bpcu is more prominent than that for 4 bpcu. Since not all reflecting elements are always activated in passive RIS-CBC-APSK, it can be seen that passive RIS-CBC-APSK slightly performs worse than RIS-star-QAM I in terms of active transmission at low $P_t$. Fortunately, thanks to the larger minimum Euclidean distance of the APSK constellations, the situation is reversed at high $P_t$. Moreover, passive RIS-CBC-APSK achieves better SER performance than RIS-star-QAM I in terms of backscatter and overall information for all $P_t$ values.

The SER performance comparison among passive RIS-CBC-APSK, RIS-star-QAM II \cite{Li_RIS_APSK}, and RIS-PSK \cite{BasarWireless} with $A\equiv 1$ and $P=M=16,32$ is illustrated in Fig.~\ref{RIS_CBC_APSK_transmitter}. In all considered schemes, the direct A-Tx to C-Rx link is ignored. As seen from Fig.~\ref{RIS_CBC_APSK_transmitter}, for both $M=16$ and $M=32$, due to the lager minimum Euclidean distance of the star-QAM constellations, RIS-star-QAM II slightly outperforms RIS-PSK. However, each ring of the star-QAM constellation in RIS-star-QAM II has the same number of points and the rings are equally spaced, which limits the error performance of RIS-star-QAM II. Fortunately, in RIS-CBC-APSK, the number of points on each ring of the APSK constellations is selected flexibly and the radii are optimized via the maximization of the minimum Euclidean distance, as described in Section II. Therefore, RIS-CBC-APSK performs significantly better than RIS-star-QAM II and RIS-PSK at the same modulation orders. Further, the gain becomes larger with increasing the modulation order.

\begin{figure*}[t]
	\centering
	\includegraphics[width=7in]{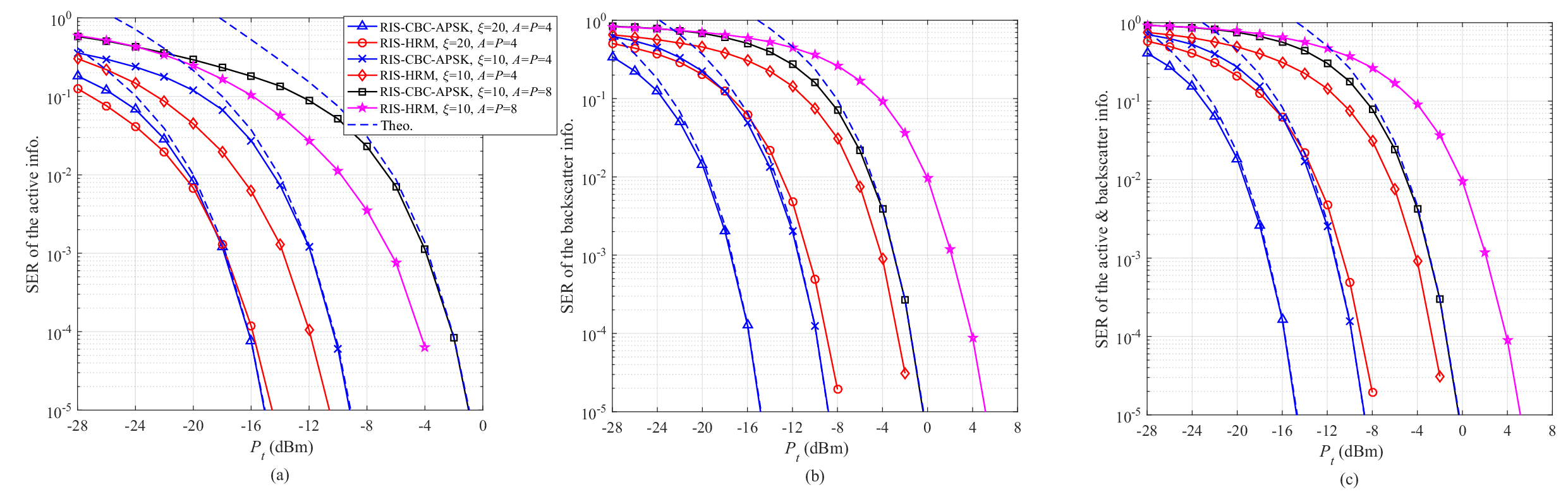}
	\caption{Performance comparison between active RIS-CBC-APSK and RIS-HRM with $A=P=4$ (4 bpcu) and $A=P=8$ (6 bpcu): (a) SER of the active information, (b) SER of the backscatter information, and (c) SER of the active and backscatter information with $\xi=10,20$.}
	\label{active_RIS_active_passive_comparison}
\end{figure*}

\begin{figure}[t]
	\centering
	\includegraphics[width=2.9in]{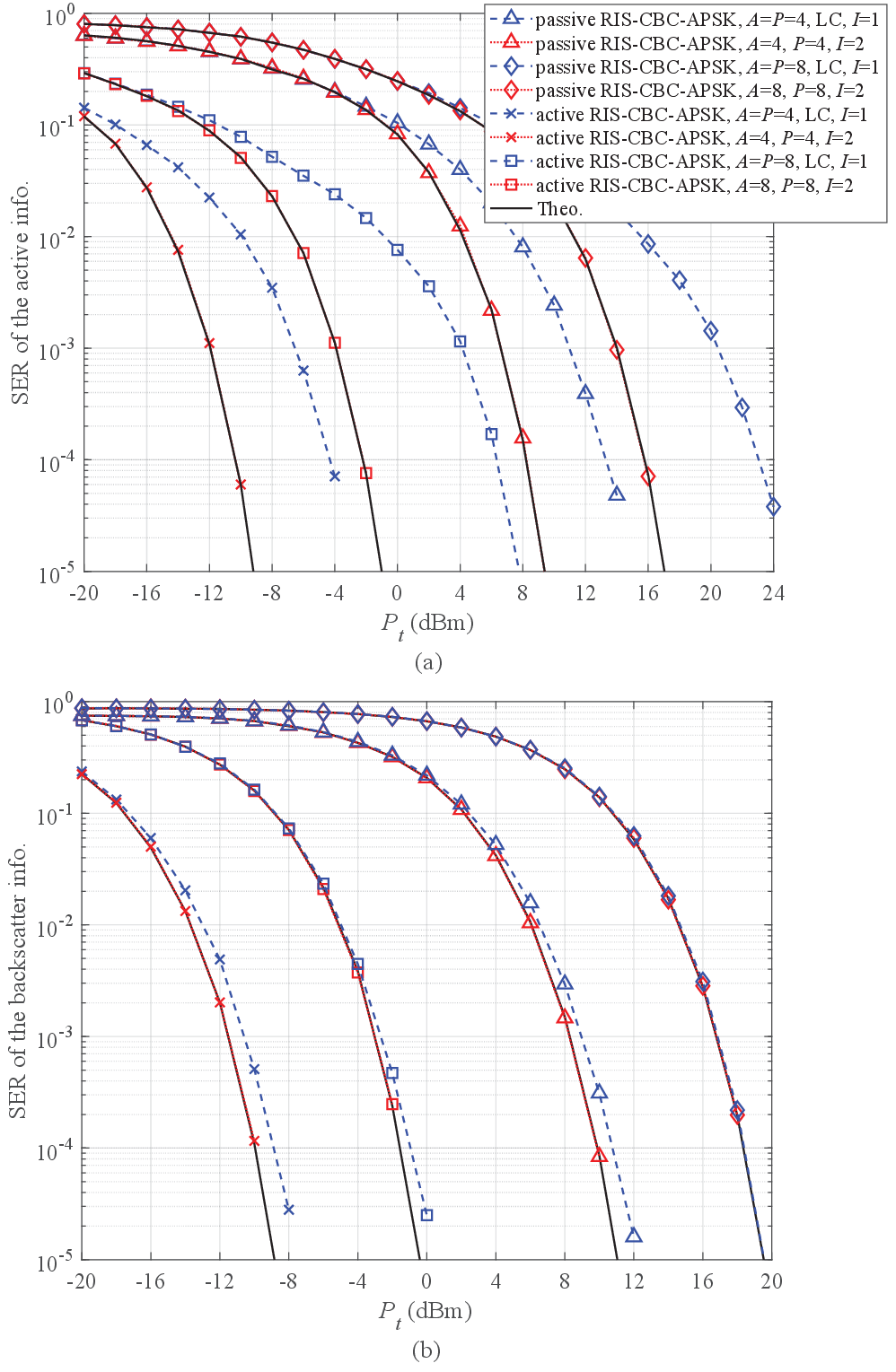}
	\caption{Performance comparison between the ML and LC detectors for passive and active RIS-CBC-APSK: (a) SER of the active information, and (b) SER of the backscatter information, where $A=P=4,8$, $\xi=10$, and $I=1,2$.}
	\label{LC_ML_comparison}
\end{figure}

\begin{figure}[t]
	\centering
	\includegraphics[width=2.6in]{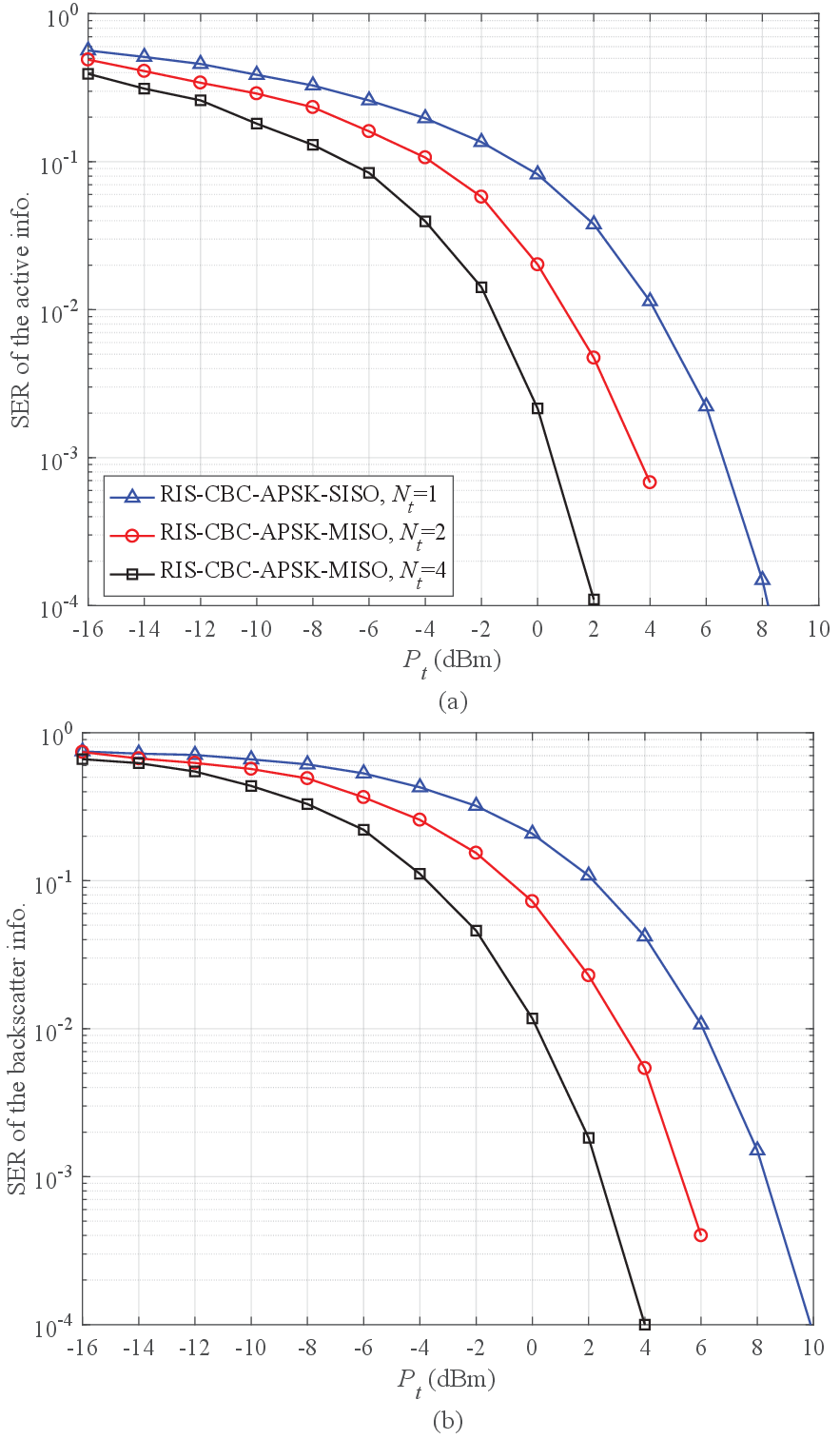}
	\caption{Performance of passive RIS-CBC-APSK in SISO and MISO scenarios, where $A=P=4$ and $N_t=1,2,4$.}
	\label{RIS_CBC_APSK_MISO}
\end{figure}

Figure~\ref{active_RIS_active_passive_comparison} depicts the SER performance comparison between active RIS-CBC-APSK and RIS-HRM \cite{Yigit_RIS_HRM} at the data rates of 4 bpcu ($A=P=4$) and 6 bpcu ($A=P=8$) with $\xi=10,20$. As shown, the theoretical curves in Figs.~\ref{active_RIS_active_passive_comparison}(a)-(c) agree with the computer simulated counterparts very well, verifying the analysis given in Section IV.B. As expected, increasing $A$ and $P$ results in the worse SER performance for both active RIS-CBC-APSK and RIS-HRM, in terms of both active and backscatter transmission, whereas increasing $\xi$ improves the performance. As seen from Fig.~\ref{active_RIS_active_passive_comparison}, compared with RIS-HRM, active RIS-CBC-APSK significantly enhances the SER performance of backscatter transmission, at the cost of performance degradation in the active transmission. Since active RIS-CBC-APSK employs APSK constellations that is optimized for simultaneously conveying both active and backscatter information, it performs better than RIS-HRM in terms of the overall SER. Hence, active RIS-CBC-APSK achieves a better trade-off between the active and backscatter transmission than RIS-HRM. Moreover, increasing $\xi$ shrinks the performance gap between active RIS-CBC-APSK and RIS-HRM in terms of the active transmission. It is also shown that active RIS-CBC-APSK outperforms RIS-HRM at medium-to-high transmit power when $\xi$ reaches the value of $20$.

The performance comparison between the ML and LC detectors for passive and active RIS-CBC-APSK, where $A=P=4,8$, $\xi=10$, and $I=1,2$ is presented in Fig.~\ref{LC_ML_comparison}. As expected, since the active RIS can amplify incident signals, all considered active RIS-CBC-APSK schemes perform better than their passive counterparts. We observe from Fig.~\ref{LC_ML_comparison} that the LC detector with $I=1$ performs worse than the ML detector for all considered RIS-CBC-APSK schemes, especially in terms of active transmission. Fortunately, the LC detector performs better with increasing values of $I$. Merely, when $I=2$, it can be observed that the LC detector almost achieves the same performance as the ML detector. This is because increasing the value of $I$ enlarges the search space of $x$. Since the LC detector with $I=2$ can be regarded as a near-optimal detector, in the case of high constellation orders, the detection complexity can be reduced by more than half without performance loss by employing the LC detector rather than the ML detector.

Finally, Fig.~\ref{RIS_CBC_APSK_MISO} illustrates the performance of passive RIS-CBC-APSK in SISO and MISO scenarios, where $A=P=4$ and $N_t=1,2,4$. It can be seen from Fig.~\ref{RIS_CBC_APSK_MISO} that, increasing the number of transmit antennas, significantly improves the SER performance of both active and backscatter transmission. This verifies the effectiveness of the AO algorithm and the feasibility of extending RIS-CBC-APSK to MISO systems. Moreover, RIS-CBC-APSK-MISO is able to unlock the large beamforming potential resulting from multiple transmit antennas. Particularly, a gain of about 3 dBm for both active and backscatter transmission is achieved by doubling the number of transmit antennas at an SER value of $10^{-3}$.

\section{Conclusions}
In this paper, we presented a new RIS-CBC scheme where a passive or an active RIS implements APSK modulation to embed backscatter information into incident unmodulated/PSK-modulated signals. The proposed RIS-based bit-mapping mechanism was designed to maximize the minimum Euclidean distance between two distinct points. Closed-form SER upper bounds were derived for both passive and active RIS-CBC-APSK with ML detection over Rician fading channels. In addition, a LC detector and an extension to MISO systems were investigated for RIS-CBC-APSK. The conducted simulation results showcased that RIS-CBC-APSK performs better than the existing RIS-NM, RIS-PSK, RIS-star-QAM, and RIS-HRM in terms of SER performance.

\end{document}